\documentclass[12pt]{iopart}

\usepackage{graphicx}
\usepackage{amsfonts, amssymb}
\usepackage{bm}

\def\im{\rmi}

\begin{document}
\title{Spontaneous-emission suppression via 
multiphoton quantum interference}
\author{J\"org Evers$^1$ and Christoph H Keitel$^{2,1}$}
\address{$^1$ Theoretische Quantendynamik, Fakult\"at f\"ur Mathematik und
Physik, Universit\"at Freiburg, Hermann-Herder-Stra{\ss}e 3, D-79104 
Freiburg, Germany}
\address{$^2$ Max-Planck-Institut f\"ur Kernphysik, Saupfercheckweg 1, D-69117 Heidelberg, Germany}
\eads{\mailto{evers@physik.uni-freiburg.de}, \mailto{keitel@mpi-hd.mpg.de}}
\begin{abstract}
The spontaneous emission is investigated for an effective atomic 
two-level system in an intense coherent field
with frequency lower than the vacuum-induced decay
width. As this additional low-frequency field is assumed to be intense,
multiphoton processes may be induced, which can be seen as alternative
transition pathways in addition to the simple spontaneous decay. 
The interplay of the various interfering transition pathways influences
the decay dynamics of the two-level system and may be used
to slow down the spontaneous decay considerably.
We derive from first principles 
an expression for the Hamiltonian
including up to three-photon processes. This Hamiltonian
is then applied to 
a quantum mechanical simulation of the decay dynamics of the
two-level system.
Finally, we discuss numerical results of this simulation
based on a rubidium atom and show that the spontaneous emission
in this system may be suppressed substantially.
\end{abstract}
\submitto{\JPB}

\section{Introduction}
While spontaneous decay is a fundamental ingredient for many
physical processes, many applications have been proposed
recently where spontaneous processes are amongst the
main limiting factors \cite{scully}. These schemes usually rely
on the persistence of either population trapped in
a specific state or of coherences on timescales
long as compared to typical atomic timescales such as
the lifetime of an atomic state. 
A well-known example for this restriction is the construction of a high frequency laser, 
where it is hard to reach a population inversion as
the spontaneous decay is considerable on high frequency
transitions. For the storage and the processing of quantum information, 
spontaneous emission is a major limitation, because
it is necessary to avoid all possible sources of decoherence in these schemes. 
The same holds for the secure information transmission using quantum effects. 

Because of the great interest in these applications,
various schemes to modify the spontaneous dynamics 
of an atomic system have been proposed
\cite{zeno,cavity,interference1,interference2,
int2b,interference3,feranchuk-1,feranchuk-2,prl,prl2,kocharovskaya}.
One ansatz is the quantum Zeno effect \cite{zeno}. According to the 
measurement postulate, a system is projected 
into one of its eigenstates upon a measurement. If the measurements 
are repeated rapidly, the system evolution may effectively be stopped,
because each measurement projects the system back into the initial
state. The technical conditions on the brevity of the pulses though 
may not always 
be easily fulfilled for every transition, especially in the free vacuum.
Another possibility is a modification of the vacuum surrounding 
the given atomic system, e.g. by an optical cavity \cite{cavity}. 
If the vacuum is modified such that the mode density at the frequency of 
an atomic transition is decreased, spontaneous processes on the given
transition may be suppressed.
Here the control of the environmental modes with cavities is rather 
challenging in reality and not suitable to all schemes.
Thirdly it is possible to find superpositions 
of more than one upper state which are stable \cite{interference1} 
or almost stable \cite{interference2,int2b} against spontaneous decay. 
The suppression of the spontaneous decay here is due to quantum 
interference effects such
as the cancellation of the amplitudes of several possible pathways between
two system states \cite{interference3}. In spite of the conceptual beauty, 
the disadvantage here 
often is the difficulty to provide convenient atomic systems which fulfill 
all conditions such as parallel transition dipole moments for 
interference to be present.
Finally, a system driven by an electromagnetic field periodic in time
is somewhat related to systems exhibiting spacial periodicity such as
crystals. Thus schemes relying on spacial periodicity which involve a 
change of the probability for incoherent
processes such as the Borrmann effect~\cite{borrmann} or the suppression 
of nuclear reactions~\cite{nuclear} may be transferred to laser-driven atomic 
systems~\cite{feranchuk-1}. Recently, this idea has also been applied to 
modify the decay of a three level system in 
V-configuration~\cite{feranchuk-2}.

In \cite{prl}, a scheme was proposed to slow down the 
spontaneous decay of the upper state population of an effective 
atomic two-level system considerably. 
This scheme makes use of an intense low-frequency laser field of 
constant frequency and intensity which is applied 
to the two-level system such that the frequency of the low-frequency field 
is lower than the decay width of the atomic transition. 
This additional field induces multiphoton transitions between the 
two atomic states under consideration,
thus allowing for alternative pathways between the two states in 
addition to spontaneous transitions.
In the literature, Hamiltonians describing
two-photon processes have been discussed to
a great extend \cite{cardimona,multiphoton}. However these mainly apply to
systems involving dipole-forbidden transitions. 
As our aim is to inhibit spontaneous decay, 
it is not sufficient to look at the spontaneous
emission on dipole-forbidden transitions as
these rates are very low naturally.
For dipole-allowed transitions, two- and four-photon pathways
vanish due to parity reasons. Therefore it is necessary to extend the analysis
to third-order processes~\cite{morethantwo}.
The leading-order corrections are five-photon processes, which we
neglect as these have a low probability in the discussed parameter range.

Thus in this paper, we derive the Hamiltonian
describing the interaction of the two-level system
with both the vacuum modes and the additional low-frequency
field mode including up to three-photon processes from first
principles by adiabatically removing the intermediate atomic
states of the multiphoton processes. 
We calculate the coupling constants which were taken as
free parameters in \cite{prl} in terms of the properties
of the atom and the laser field. Furthermore, the Hamiltonian
includes Stark shift effects not discussed in \cite{prl}, which however
will turn out to be of no importance for the suppression scheme.
This Hamiltonian is then applied to a quantum mechanical 
simulation of the decay dynamics of the two-level system. 
As it will turn out, the spontaneous decay may be slowed down considerably 
due to the additional transition pathways induced
by the intense low-frequency field.
In the final part, this is demonstrated using numerical results of 
this simulation based on rubidium atoms.

\section{Derivation of the multiphoton Hamiltonian}
The derivation starts from the usual Hamiltonian for an
 atomic system coupled to quantized electromagnetic
fields \cite{scully}. The atomic system consists of two atomic 
states (typically of opposite parity) which are considered as 
the ground state $|1\rangle$ and the excited state $|2\rangle$ of
an two-level atom, which we will denote as the effective
two-level system throughout this paper. 
To model possible multiphoton transitions between the
two effective atomic levels, unspecified auxiliary states 
$|j\rangle$ ($j=3,4,\dots$) are required,
which will be adiabatically eliminated throughout the 
analysis.
\begin{figure}[b]
\begin{center}
\includegraphics[height=4.5cm]{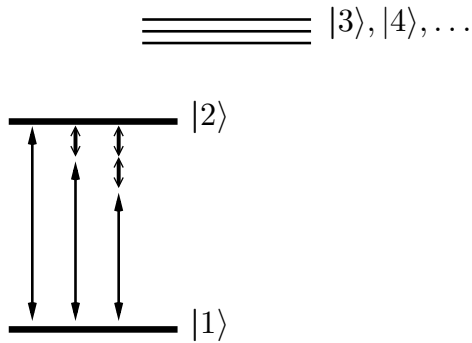}
\end{center}
\caption{The two-level atom and the auxiliary states  
$|3\rangle, |4\rangle, \dots$ which will be 
adiabatically removed in the derivation of the multiphoton Hamiltonian.
The auxiliary states are show schematically; their relative position 
is not restricted to be above the two main atomic states.
After the adiabatic elimination, the effective two-level system exhibits
multiphoton processes  as symbolized in the figure. Here, the 
short arrows depict interactions with the intense low-frequency field and
the long arrows interactions with the vacuum field. 
}
\end{figure}
These states might even be part of the continuum; the only condition 
on them is that they are sufficiently far away from the two main 
atomic states as will be explained in more detail below.
Since there are no isolated two-level atomic systems, the presence of 
this states is always guaranteed.  
The auxiliary states are necessary as in starting from
first principles, i.e. the interaction Hamiltonian $H_I = -\vec{d}\vec{E}$,
we only allow for one-photon transitions between the two effective atomic
states. Multiphoton processes are then introduced by considering
series of one-photon processes from one of the two effective states
to the other effective state via one or more of the auxiliary states,
which within the adiabatic approximation occur simultaneously.
The result will be a Hamiltonian involving the states of the 
effective two-level system only, which will turn out to 
be a generalization of the one used in \cite{prl}.

The applied field consists of the low frequency field of 
frequency $\bar{\omega}$
and the vacuum field modes. 
The following analysis is similar to the one in \cite{cardimona},
and the notation is close to  \cite{cardimona, prl}. 
The system Hamiltonian is given by
\begin{eqnarray}
\fl H = H_0+H_I \, ,\\
\fl H_0 = \hbar \omega_1 |1\rangle\langle 1| + \hbar \omega_2 
       |2\rangle\langle 2|\: 
      + \sum_{j\notin \{1,2\}} \hbar \omega_j |j\rangle\langle j| 
       + \hbar \bar{\omega}b^\dagger b + \hbar \sum_k \omega_k 
       a_k^\dagger a_k \: ,\\
\fl H_I = -\vec{d}\vec{E}\: .
\end{eqnarray}
Here, $\hbar \omega_1, \hbar \omega_2$ and $\hbar \omega_j$ ($j=3,4,\dots$) 
are the energies of state $|1\rangle, |2\rangle$ and auxiliary state
$|j\rangle$, respectively.
$k$ is a multi index over all polarization modes 
and vacuum frequencies $\omega_k$.
Throughout the analysis, we denote sums over all auxiliary states 
excluding the two effective atomic states $|1\rangle, |2\rangle$ 
by the sum subscript $j\notin \{1,2\}$, where $j$ is the summation index.
$b$ ($b^\dagger$) is a low-frequency field annihilation (creation) operator,
$a_k$ $(a_k^\dagger$) annihilates (creates)
a vacuum photon of mode $k$, $\vec{d}$ is the total atomic dipole operator, 
and $\vec{E}$ is the electric field given by
\begin{eqnarray}
\vec{E} = \im \left ( \mathcal{E}(\vec{e}b - \vec{e}^*b^\dagger) 
 + \sum_k \mathcal{E}_k(\vec{e}_k a_k 
 - \vec{e}_k ^*a_k^\dagger) \right )\: . \label{field}
\end{eqnarray}
The $\vec{e}, \vec{e}_k$ are polarization vectors, while the 
$\mathcal{E}, \mathcal{E}_k$ are field amplitudes.
The frequency corresponding to the low-frequency field operators 
$b, b^\dagger$ is assumed
to be different from the relevant frequencies for the vacuum photons, which are
close to the atomic transition frequency of the effective two-level system,
such that the low-frequency field operators commute with the relevant 
vacuum field operators.
In the following analysis, we make use of the approximation that 
no multiphoton transitions involving two or more spontaneous photons 
with {\it different} frequencies are included. As in our system
spontaneous one-photon transitions are possible, transitions
involving more than one spontaneous photon are highly suppressed; 
see the processes considered and observed in \cite{mossberg}.
However we include multiple spontaneous photons with the {\it same} frequency
to account for Stark shifts, which arise from interactions of the
atom with the electromagnetical field as a sequence of an
absorption and an emissions (or vice versa) which does not change the state of
the atomic system.
Also we consider all possible combinations of interactions with
the low-frequency field as it is intense.

The interaction part of the Hamiltonian may be expanded as follows:
\begin{eqnarray}
\fl H_I= -\vec{d}\vec{E} = - \sum_{n,m} |n\rangle\langle n|\vec{d}
     |m\rangle\langle m| \vec{E} 
= - \sum_{n,m} \vec{d}_{nm}\sigma_{nm}\vec{E} \nonumber\\
\lo= - \left ( \vec{d}_{12}\sigma_{12} + \vec{d}_{21}\sigma_{21} \right ) 
     \vec{E}  
  - \sum_{j\notin \{1,2\}} \left ( \vec{d}_{1j}\sigma_{1j} +
     \vec{d}_{j1}\sigma_{j1}  
   + \vec{d}_{2j}\sigma_{2j} 
   + \vec{d}_{j2}\sigma_{j2}\right ) \vec{E} \nonumber \\  
   - \sum_{j,j'\notin \{1,2\}}  \vec{d}_{jj'}\sigma_{jj'}  \vec{E}\: .
\label{order3}
\end{eqnarray}
$\sigma_{nm} = |n\rangle\langle m|$ is the atomic transition operator, 
$\vec{d}_{nm} = \langle n |\vec{d} | m\rangle$ is the dipole moment
corresponding to transition $n\leftrightarrow m$.
Some of these dipole moments may be zero due to symmetry
reasons, however as this depends on the configuration of the auxiliary
states, we keep this most general form of the Hamiltonian. A specific
example for this will be given in section~\ref{rubidium} where we discuss
rubidium as a model system.
In this expansion, the first part in equation~(\ref{order3}) corresponds to direct 
transitions between the ground and the excited state by dipole coupling to one 
of the electric field components (i.e. one-photon transitions). The second
addend couples both the ground and the excited state to one of the auxiliary 
states. After an adiabatic elimination of these intermediate
states, this part will turn out to represent multiphoton transitions. 
Obviously n-photon transitions with $n>2$ require transitions amongst the
auxiliary states; thus these effects are accounted for by the last part 
in equation~(\ref{order3}).
These last two parts together with the intense low-frequency field
lead to a system evolution which qualitatively differs from the
well-known Mollow-type evolution for two-level systems resonantly
driven by a laser field \cite{mollow}.
It is important to note that a higher order multiphoton treatment based on the 
above Hamiltonians involves operator ordering issues, as in the following
analysis the atomic transition operators involving auxiliary states will
be represented by expressions containing photon operators. If these issues
are not taken into account, the resulting Hamiltonians are not self-adjoint.
Thus strictly speaking, expressions like 
\begin{equation}
(\vec{d}_{1j}\sigma_{1j} + \vec{d}_{j1}\sigma_{j1})(\vec{e}b 
  - \vec{e}^*b^\dagger)
\end{equation}
in the above Hamiltonians are understood to be read as 
\begin{equation}
\vec{d}_{j1}\sigma_{j1} (\vec{e} b - \vec{e}^*b^\dagger) 
+ (\vec{e}b - \vec{e}^*b^\dagger)\: \vec{d}_{1j}\sigma_{1j}\: .
\end{equation}
As the above equation equation~(\ref{order3}) is difficult to solve exactly,
we use a perturbative approach which consists of an expansion
in the number of interactions with the auxiliary levels.
The lowest order Hamiltonian thus does not include any interactions
with the auxiliary levels and may therefore be written as 
\begin{equation}
H_I^{(1-photon)} = - \left ( \vec{d}_{12}  \vec{E} \sigma_{12} 
     + \vec{d}_{21}\sigma_{21} \vec{E} \right )\: .
\end{equation}
This Hamiltonian simply describes the coupling of the vacuum- and the 
low-frequency field to the effective two-level atom and does not
contain contributions of the auxiliary states.
To account for two-photon processes, we have to consider transitions 
from the ground or the excited state into and out of the auxiliary states, but
no transitions amongst the auxiliary states. For this we introduce
the transition operators $\sigma_{1j}^{(0)}, \sigma_{2j}^{(0)}$
and their conjugates, where the super index $(0)$ denotes the lowest order 
approximation whose equations of motion 
do not include any contribution of the auxiliary states. 
Only the operators involving the auxiliary states are approximated, as the 
operators connecting only the ground and the excited state contain the 
main evolution
of the system and therefore have to be taken into account to all orders.
Thus the two-photon Hamiltonian is given by
\begin{eqnarray}
 H_I^{(2-photon)} 
=& - \left ( \vec{d}_{12}\vec{E}\sigma_{12} + \vec{d}_{21}\sigma_{21}\vec{E} \right )   
- \sum_{j\notin \{1,2\}} \Bigl ( \vec{d}_{1j}\vec{E}\sigma_{1j}^{(0)} + \vec{d}_{j1}\sigma_{j1}^{(0)} \vec{E} \nonumber \\
 &+ \vec{d}_{2j}\vec{E}\sigma_{2j}^{(0)} + \vec{d}_{j2}\sigma_{j2}^{(0)}\vec{E}\Bigr )  \: .\label{h2}
\end{eqnarray}
As it will be shown in section~\ref{transops}, the transition 
operators $\sigma_{1j}^{(0)}, \sigma_{2j}^{(0)}$ and their conjugates
consist of sums of products of a transition operator $\sigma_{ij}$ 
($i,j\in \{1,2\}$) and a photon annihilation
or creation operator. Thus it may easily be seen that contributions 
to the Hamiltonian in equation~(\ref{h2}) containing transition operators to one of
the auxiliary states describe two-photon processes.
Including up to three-photon processes, a transition from one of the two
effective atomic levels to one of the auxiliary levels may
be followed by a transition to another auxiliary state. Thus
the three-photon Hamiltonian may be expressed in terms of the 
first order transition operators $\sigma_{1j}^{(1)}, \sigma_{2j}^{(1)}$ 
and their conjugates, whose equations of motion contain either 
lowest-order transition operators to the auxiliary state or 
transitions between the two effective atomic states. Also,
transition operators between two auxiliary states are possible in
lowest order $\sigma_{jj'}^{(0)}$ ($j,j' \notin \{1,2\}$), i.e. with
no reference to the auxiliary states in the equations of motion.
This yields
\begin{eqnarray}
\fl H_I^{(3-photon)} = - \left ( \vec{d}_{12}\vec{E}\sigma_{12} 
   + \vec{d}_{21}\sigma_{21} \vec{E}\right )  
 - \sum_{j,j'\notin \{1,2\}}  \vec{d}_{jj'}\sigma_{jj'}^{(0)} \vec{E}
  \nonumber\\
\lo - \sum_{j\notin \{1,2\}} \Bigl ( \vec{d}_{1j}\vec{E}\sigma_{1j}^{(1)} 
   + \vec{d}_{j1}\sigma_{j1}^{(1)}\vec{E} 
   + \vec{d}_{2j}\vec{E}\sigma_{2j}^{(1)} 
   + \vec{d}_{j2}\sigma_{j2}^{(1)}\vec{E}\Bigr )    
  \label{h3}
\end{eqnarray}
as the three-photon Hamiltonian. Here, the operator ordering was not 
applied to the
term containing a double sum over $j,j'$ for notational simplicity as it will
turn out to be irrelevant for the present analysis.
For higher order processes, $H_I^{(n-photon)}$ 
with $n>3$ may be obtained similarly.
Introducing the coupling constants
\begin{equation}
\lambda_{ij} = - \frac{\im\: \mathcal{E} \: \vec{d}_{ij} \: \vec{e}}{\hbar}\:,
\qquad 
\lambda_{ijk} = - \frac{\im\: \mathcal{E}_k  \: \vec{d}_{ij} \:
                  \vec{e}_k}{\hbar}  \label{coupling}
\end{equation}
where the transition dipole moments are assumed to be real,
the Hamiltonians may be written as
\begin{eqnarray}
\fl H_I^{(1-photon)} = \hbar \: \Bigl \{ \lambda_{12}\: b 
  + \lambda_{12}^* \: b^\dagger  
   + \sum_k \left (  \lambda_{12k}\: a_k 
  + \lambda_{12k}^* \: a_k^\dagger \right )  \Bigr \}\:\sigma_{12} 
  + \textrm{ h.c.}\: , \\
\fl H_I^{(2-photon)} = H_I^{(1-photon)}
 \nonumber \\
\lo  + \hbar \sum_{m=1}^{2}\sum_{j\notin \{1,2\}} \Bigl \{ 
  \lambda_{mj}\: b + \lambda_{mj}^* \: b^\dagger   
  + \sum_k  \left ( \lambda_{mjk}\: a_k 
  + \lambda_{mjk}^* \: a_k^\dagger \right ) \Bigr \}\: \sigma_{mj}^{(0)}  
  + \textrm{ h.c.}\: , \label{twophot} \\
\fl H_I^{(3-photon)} = H_I^{(1-photon)} \nonumber \\
\lo  + \hbar \sum_{m=1}^{2}\sum_{j\notin \{1,2\}} \Bigl \{ 
\lambda_{mj}\: b + \lambda_{mj}^* \: b^\dagger  
  + \sum_k  \left ( \lambda_{mjk}\: a_k 
  + \lambda_{mjk}^* \: a_k^\dagger \right ) \Bigr \}\: \sigma_{mj}^{(1)}  
 \nonumber \\
\lo +\hbar \sum_{j,j'\notin \{1,2\}}  \Bigl \{ \lambda_{jj'}\: b 
+ \lambda_{jj'}^* \: b^\dagger  
   + \sum_k  \left ( \lambda_{jj'k}\: a_k 
+ \lambda_{jj'k}^* \: a_k^\dagger \right ) \Bigr \}\: \sigma_{jj'}^{(0)} 
+ \textrm{ h.c.}\: . \label{threephot}
\end{eqnarray}

\subsection{\label{transops}Atomic transition operators}
\subsubsection{Equations of motion}
The equations of motion for the various operators involved may be obtained 
using the Heisenberg equation
\begin{equation}
 \frac{d}{dt} O = \frac{\im}{\hbar} [H, O]
\end{equation}
where $O$ is an operator in the Heisenberg picture.
For the transition operator $\sigma_{ij}$ with $i=1, j\notin \{1,2\}$ 
- which is one of the elements that occurs in 
the above Hamiltonians in equations~(\ref{twophot}, \ref{threephot}) - this may 
be expanded as follows to the
lowest two orders in the interaction with the auxiliary states:
\begin{eqnarray}
\fl \dot{\sigma}_{1j}^{(0)} = \im \: \omega_{1j}\sigma_{1j}^{(0)} + 
       \frac{\im}{\hbar} \left ( \vec{d}_{j1}\sigma_{11} 
       + \vec{d}_{j2}\sigma_{12}  \right ) \vec{E}\: , \label{s1j_zero} \\
\fl \dot{\sigma}_{1j}^{(1)} =\im \: \omega_{1j}\sigma_{1j}^{(1)} 
  + \frac{\im}{\hbar} \left ( \vec{d}_{j1}\sigma_{11} 
                            + \vec{d}_{j2}\sigma_{12} \right ) \vec{E}
 - \frac{\im}{\hbar} \left ( \vec{d}_{21}\vec{E}\sigma_{2j}^{(0)} 
   - \sum_{n\notin \{1,2\}}  \vec{d}_{jn}\sigma_{1n}^{(0)} \vec{E} \right ) \, ,
    \label{s1j_one}
\end{eqnarray}
where $\omega_{nm} = \omega_n - \omega_m$ ($n,m \in \mathbb{N}$).
The zeroth order equation  does not include references to the auxiliary 
states, while the first
order includes operators connecting one of the main atomic states with 
an auxiliary state
in zeroth order. 
Note that for simplicity we omitted an addend 
containing the transition operator $\sigma_{nj}^{(0)}$ in the 
first-order equation, as it will turn out to be zero (see 
equation (\ref{aux-pop})).
Higher order expressions may be obtained similarly.
For $i=2,j\notin \{1,2\}$, we have:
\begin{eqnarray}
\fl \dot{\sigma}_{2j}^{(0)} = \im \: \omega_{2j}\sigma_{2j}^{(0)} 
+ \frac{\im}{\hbar} \left ( \vec{d}_{j1}\sigma_{21} 
       + \vec{d}_{j2}\sigma_{22}  \right ) \vec{E}\: , \\
\fl \dot{\sigma}_{2j}^{(1)} = \im \: \omega_{2j}\sigma_{2j}^{(1)}
 + \frac{\im}{\hbar} \left ( \vec{d}_{j1}\sigma_{21} 
       + \vec{d}_{j2}\sigma_{22} \right ) \vec{E}  
 - \frac{\im}{\hbar} \left ( \vec{d}_{12}\vec{E} \sigma_{1j}^{(0)} 
   - \sum_{n\notin \{1,2\}}  
    \vec{d}_{jn}\sigma_{2n}^{(0)} \vec{E} \right )\: .  \label{s2j_one}
\end{eqnarray}
For $i,j\notin \{1,2\}$ the corresponding expression in lowest order 
simply becomes
\begin{eqnarray}
\dot{\sigma}_{ij}^{(0)} = \im \: \omega_{ij}\sigma_{ij}^{(0)}\: .
\label{eom_ij}
\end{eqnarray}

\subsubsection{\label{first-order}First-order transition operators}
The basic tool to eliminate the auxiliary states from the equation of 
motion is the adiabatic integration. For this, the equations of motion are 
written in terms of slowly varying operators which are denoted by the 
corresponding symbol with a tilde and which may be seen as an interaction 
picture representation of the operator:
\begin{eqnarray}
\tilde{\sigma}_{12} = \sigma_{12} \: e^{\im (\bar{\omega}+\omega_k)t}, \qquad 
  & \tilde{\sigma}_{1j} = \sigma_{1j} \: e^{-\im  \omega_{1j}t}, \nonumber \\
\tilde{\sigma}_{ii} = \sigma_{ii}, 
  & \tilde{\sigma}_{2j} = \sigma_{2j} \: e^{-\im  \omega_{2j}t}, \nonumber \\
\tilde{b} = b \: e^{\im  \bar{\omega} t}, 
  & \tilde{a}_k = a_k \: e^{\im  \omega_k t}. \label{trans}
\end{eqnarray}
The adiabatic approximations with superscripts $(0), (1)$ are transferred
as their full counterparts. The definition of the slowly varying operators
is chosen as in \cite{cardimona} to allow for a comparison of the results.
This for example yields using  equations~(\ref{field}), (\ref{coupling}) 
and (\ref{s1j_zero}) 
\begin{eqnarray}
\fl \dot{\tilde{\sigma}}_{1j}^{(0)} = - \im \: \lambda_{j1} \:
 \tilde{\sigma}_{11} \: \tilde{b} \: e^{-\im (\omega_{1j}+\bar{\omega}) t}
-\im \:  \lambda_{j1}^* \: \tilde{\sigma}_{11} \: \tilde{b}^\dagger  \:
 e^{-\im (\omega_{1j}-\bar{\omega}) t} 
- \im \:  \lambda_{j1k} \: \tilde{\sigma}_{11} \: \tilde{a}_k \: 
e^{-\im (\omega_{1j}+\omega_k) t} \nonumber \\
\fl\qquad 
- \im \:  \lambda_{j1k}^* \: \tilde{\sigma}_{11} \: \tilde{a}^\dagger _k \:
 e^{-\im (\omega_{1j}-\omega_k) t} 
- \im \: \lambda_{j2} \: \tilde{\sigma}_{12} \: \tilde{b} \: 
e^{-\im (\omega_{1j}+2\bar{\omega}+\omega_k) t} 
- \im \:  \lambda_{j2}^* \: \tilde{\sigma}_{12} \: \tilde{b}^\dagger  \:
 e^{-\im (\omega_{1j}+\omega_k) t}  \nonumber \\
\fl\qquad
 - \im \: \lambda_{j2k} \: \tilde{\sigma}_{12} \: \tilde{a}_k \: 
e^{-\im (\omega_{1j}+2\omega_k+\bar{\omega}) t} 
 - \im \:  \lambda_{j2k}^* \: \tilde{\sigma}_{12} \: \tilde{a}^\dagger _k \:
 e^{-\im (\omega_{1j}+\bar{\omega}) t} \: . \label{adiab-example}
\end{eqnarray}
Now we integrate over time, using the partial integration rule. For
simplicity, only the first addend of equation~(\ref{adiab-example}) is shown to
illustrate the basic idea of adiabatic integration:
\begin{eqnarray}
\tilde{\sigma}_{1j}^{(0)} &=& - \im \: \int \lambda_{j1} \: 
\tilde{\sigma}_{11} \: \tilde{b} \: e^{-\im (\omega_{1j}+\bar{\omega}) t} dt 
+ (\textrm{other addends}) \nonumber\\
&=& \frac{\lambda_{j1} \: \tilde{\sigma}_{11} \: \tilde{b}}
{ (\omega_{1j}+\bar{\omega})}\: e^{- \im (\omega_{1j}+\bar{\omega}) t}
- \int \frac{\lambda_{j1}  e^{-\im (\omega_{1j}+\bar{\omega}) t}}
{ (\omega_{1j}+\bar{\omega})}\: \left \{ \frac{d}{dt}
\left ( \tilde{\sigma}_{11} \: \tilde{b}\right )\right \}  \:  dt 
\nonumber\\
&& + (\textrm{other addends}) \: .\label{partint}
\end{eqnarray}
In the adiabatic integration one now assumes that the  
time evolution of the slowly changing operators
\begin{equation}
\frac{d}{dt}\left ( \tilde{\sigma}_{11} \: \tilde{b}\right )
\end{equation}
(which is typically of the order of the atomic decay rate 
or the Rabi frequencies involved) is low as compared to the 
oscillation of the exponential function. This is fulfilled if 
the Rabi frequencies involved are not too large
and if the intermediate states $|j\rangle$ ($j\notin \{1,2\}$) are 
sufficiently far away from the atomic ground and excited state,
which we assume in the following.
Therefore the integral on the right hand side of equation (\ref{partint}) 
may be dropped, yielding
using (\ref{trans}) and including all addends:
\begin{eqnarray}
\fl \sigma_{1j}^{(0)} =
\frac{\lambda_{j1} \: \tilde{\sigma}_{11} \: \tilde{b}}
{(\omega_{1j}+\bar{\omega})}\: e^{-\im\bar{\omega} t}  
+\frac{\lambda_{j1}^* \: \tilde{\sigma}_{11} \: \tilde{b}^\dagger }
{(\omega_{1j}-\bar{\omega})}\: e^{\im\bar{\omega} t} 
+\frac{\lambda_{j2} \: \tilde{\sigma}_{12} \: \tilde{b}}
{(\omega_{1j}+2\bar{\omega}+\omega_k)}\: e^{-\im(2\bar{\omega}+\omega_k) t}  
\nonumber \\
\lo+ \frac{\lambda_{j2}^* \: \tilde{\sigma}_{12} \: \tilde{b}^\dagger }
{(\omega_{1j}+\omega_k)}\: e^{-\im\omega_k t} 
+\frac{\lambda_{j2k} \: \tilde{\sigma}_{12} \: \tilde{a}_k}
{(\omega_{1j}+\bar{\omega}+2\omega_k)}\: e^{-\im(2\omega_k+\bar{\omega}) t}  
+\frac{\lambda_{j2k}^* \: \tilde{\sigma}_{12} \: \tilde{a}_k^\dagger }
{(\omega_{1j}+\bar{\omega})}\: e^{-\im\bar{\omega} t}
 \nonumber \\ 
\lo+ \frac{\lambda_{j1k} \: \tilde{\sigma}_{11} \: \tilde{a}_k}
{(\omega_{1j}+\omega_k)}\: e^{-\im\omega_k t}  
+\frac{\lambda_{j1k}^* \: \tilde{\sigma}_{11} \: \tilde{a}_k^\dagger }
{(\omega_{1j}-\omega_k)}\: e^{\im\omega_k t}  
. \label{s1j_one_int}
\end{eqnarray}
as the lowest order adiabatic approximation for this atomic 
transition operator. 
A similar calculation yields
\begin{eqnarray}
 \sigma_{2j}^{(0)} =
\frac{\lambda_{j2} \: \tilde{\sigma}_{22} \: \tilde{b}}
{(\omega_{2j}+\bar{\omega})}\: e^{-\im\bar{\omega} t}  
+\frac{\lambda_{j2}^* \: \tilde{\sigma}_{22} \: \tilde{b}^\dagger }
{(\omega_{2j}-\bar{\omega})}\: e^{\im\bar{\omega} t} 
+\frac{\lambda_{j2k} \: \tilde{\sigma}_{22} \: \tilde{a}_k}
{(\omega_{2j}+\omega_k)}\: e^{-\im\omega_k t}  
\nonumber \\
+\frac{\lambda_{j2k}^* \: \tilde{\sigma}_{22} \: \tilde{a}_k^\dagger }
{\im(\omega_{2j}-\omega_k)}\: e^{\im\omega_k t}   
 +\frac{\lambda_{j1} \: \tilde{\sigma}_{21} \: \tilde{b}}
{(\omega_{2j}-\omega_k)}\: e^{\im\omega_k t}  
+\frac{\lambda_{j1}^* \: \tilde{\sigma}_{21} \: \tilde{b}^\dagger }
{(\omega_{2j}-\omega_k-2\bar{\omega})}\: e^{\im(\omega_k+2\bar{\omega}) t}
  \nonumber \\
+\frac{\lambda_{j1k} \: \tilde{\sigma}_{21} \: \tilde{a}_k}
{(\omega_{2j}-\bar{\omega})}\: e^{\im\bar{\omega} t}  
+\frac{\lambda_{j1k}^* \: \tilde{\sigma}_{21} \: \tilde{a}_k^\dagger }
{(\omega_{2j}-2\omega_k-\bar{\omega})}\: e^{\im(2\omega_k+\bar{\omega}) t} .
 \label{s2j_one_int}
\end{eqnarray}
From equation~(\ref{eom_ij}), we obtain for $i,j\notin \{1,2\}$ in the 
slowly varying operator frame:
\begin{eqnarray}
\dot{\tilde{\sigma}}_{ij}^{(0)} = 0\: .
\end{eqnarray}
As the auxiliary states are far from resonance with the applied frequencies, 
in lowest order 
of the adiabatic approximation we thus have \cite{cardimona}
\begin{eqnarray}
\tilde{\sigma}_{ij}^{(0)} = 0 = \sigma_{ij}^{(0)}\:  \label{aux-pop}
\end{eqnarray}
for $i,j\notin \{1,2\}$.
Another way of seeing this is that due to the vanishing time derivative, 
in this order of approximation
the populations of the auxiliary states remain constant. As these 
populations are zero if the population
initially is distributed over the ground and the excited state, they are 
empty at all times.

These results are sufficient to calculate the effective two-photon 
Hamiltonian equation~(\ref{h2}) and to obtain
the first order expressions for the transition operators in 
equations~(\ref{s1j_one}) and (\ref{s2j_one}).
As the expressions in equations~(\ref{s1j_one_int}) and (\ref{s2j_one_int}) 
only contain transition operators involving
the ground and the excited state, the elimination of the auxiliary states 
both from the two-photon Hamiltonian
and from the first order expressions for the transition operators is 
obvious. Note however that the 
dipole moments connecting the ground and the excited atomic state to 
the auxiliary states remain in 
the equations influencing the various coefficients or coupling parameters.

\subsubsection{\label{higher-order}Higher-order transition operators}
To calculate the first order transition operators, we 
we insert the lowest-order transition operators
equations~(\ref{s1j_one_int}) and (\ref{s2j_one_int}) and
their conjugates in the equations of motion for the first order operators
equations~(\ref{s1j_one}) and (\ref{s2j_one}).  Then the equations are
transferred to the slowly changing operator picture using the transformation
rules equations~(\ref{trans}).
The resulting expression for $\dot{\tilde{\sigma}}_{1j}^{(1)}$,
$\dot{\tilde{\sigma}}_{2j}^{(1)}$ 
and their conjugates may again be adiabatically integrated yielding
$\sigma_{1j}^{(1)}, \sigma_{j1}^{(1)}, \sigma_{2j}^{(1)}$ and
$\sigma_{j2}^{(1)}$. In addition to the corresponding operator of zeroth 
order, each of these operators contains 56 addends as  first order
contribution, 28 of which depend on a sum over auxiliary intermediate 
atomic states $|n\rangle$.
For example, we have
\begin{equation}
\sigma_{1j}^{(1)} = \sigma_{1j}^{(0)} + A_{1j}^{22} + A_{1j}^{21} 
+ \sum_{n\notin \{1,2\}} \left ( A_{1j}^{11}(n) + A_{1j}^{12}(n) \right ) \: .
\end{equation}
The operators $A_{ij}$, which are proportional to $\sigma_{ij}$,
and the other first order transition operators may be found in 
\ref{anhang-trans-op}.


\subsection{Effective Hamiltonians}
In this section, we will use the transition operators derived in the 
last section to give an explicit representation of the two-photon 
and the three-photon Hamiltonian in
equations~(\ref{h2}) and (\ref{h3}). The resulting Hamiltonian will turn out to
be equivalent to the Hamiltonian used in \cite{prl}. However here we 
obtain expressions for the coupling constants $g_k, \bar{g}_i$ introduced 
as free parameters in the Hamiltonian in \cite{prl} and extend the analysis 
to include Stark shift contributions to the Hamiltonian.
The role of the Stark shifts for the decay dynamics of the effective 
two-level atom will be discussed in section~\ref{decay}.

\subsubsection{\label{sec-two-photon}Two-photon Hamiltonian}
Using the rotating-wave approximation, we drop all terms oscillating 
with frequencies of the order of $\omega_k$ or higher in the two-photon 
Hamiltonian in equation~(\ref{h2}). 
Thus we neglect counter-rotating interactions with the vacuum field,
but do not apply the rotating-wave approximation to the 
interaction of the atom with the low-frequency field.
Transferring the resulting expression back 
to the Heisenberg picture adopted in the initial equations
equation~(\ref{order3}) using the transformation rules in equations~(\ref{trans}), 
we obtain
\begin{eqnarray}
\fl H_I^{(2-photon)} = \hbar \: \sum_k \: \Bigl \{ 
  \alpha_0 \: a_k \: \sigma_{21}
  +\alpha_1\:b\:a_k\:\sigma_{21} 
 +\alpha_2\:b^\dagger\:a_k\:\sigma_{21} + \textrm{  h.c.}
  \Bigr \} 
\nonumber \\ 
\fl\quad + \hbar \: \sum_k \: ( \alpha_3\:a_k^\dagger \: a_k + \alpha_4\: 
    a_k \: a_k^\dagger) \: \sigma_{11} 
 +\: \hbar \: (\alpha_5\: b^\dagger \: b + \alpha_6\: b \: b^\dagger
 +\alpha_{11}\:b \: b +  \:\alpha_{11}^* \:b^\dagger \: b^\dagger
   ) \:\sigma_{11}
\nonumber \\
\fl \quad  + \hbar \: \sum_k \:( \alpha_7\:a_k^\dagger \: a_k + \alpha_8\: 
     a_k \: a_k^\dagger) \: \sigma_{22} 
 + \: \hbar \: (\alpha_9\: b^\dagger \: b + \alpha_{10}\: b \: b^\dagger
  + \alpha_{12}\:b \: b +  \: \alpha_{12}^* \:b^\dagger \: b^\dagger
  ) \:\sigma_{22}  \, ,
\label{ham2_ad} 
\end{eqnarray}
where the coefficients are given by $\alpha_0 = \lambda_{12k}$,
\begin{eqnarray}
\fl \alpha_1 = \sum_{j\notin \{1,2\}} \Biggl (
 \frac{\lambda _{2j}\lambda _{j1k}}{(\Delta -\Delta_j ) } 
- \frac{\lambda _{2j}\lambda _{j1k}}{\Delta_j } 
 - \frac{\lambda _{2jk}\lambda _{j1}}{(\Delta_j +\omega_k -\bar{\omega} ) }
+ \frac{\lambda _{2jk}\lambda _{j1}}{(\Delta -\Delta_j -\omega_k 
+\bar{\omega} ) }
\Biggr )\: , \label{alpha1}\\
\fl \alpha_2 = \sum_{j\notin \{1,2\}} \Biggl (
 \frac{\lambda _{j1k}\lambda^* _{2j}}{(\Delta -\Delta_j ) } 
- \frac{\lambda _{j1k}\lambda^* _{2j}}{(\Delta_j -2 \bar{\omega} ) } 
+ \frac{\lambda _{2jk}\lambda^* _{j1}}{(\Delta -\Delta_j 
- \omega_k -\bar{\omega} ) } 
- \frac{\lambda _{2jk}\lambda^* _{j1}}{(\Delta_j +\omega_k -\bar{\omega} ) } 
\Biggr )\: , 
\end{eqnarray}
\begin{eqnarray}
\fl \alpha_3 = \sum_{j\notin \{1,2\}}  
- \frac{2\lambda _{j1k}\lambda^* _{1jk}}{\Delta_j } 
\: ,  
& \fl \alpha_4 = \sum_{j\notin \{1,2\}} 
- \frac{2\lambda _{1jk}\lambda^* _{j1k}}{(\Delta_j +2 \omega_k ) } 
\: ,\\
\fl \alpha_5 = \sum_{j\notin \{1,2\}} 
- \frac{2\lambda _{j1}\lambda^* _{1j}}{(\Delta_j +\omega_k -\bar{\omega} ) } 
\: , 
& \fl \alpha_6 = \sum_{j\notin \{1,2\}} 
- \frac{2\lambda _{1j}\lambda^* _{j1}}{(\Delta_j +\omega_k +\bar{\omega} ) } 
\: ,\\
\fl \alpha_7 = \sum_{j\notin \{1,2\}} 
 \frac{2\lambda _{j2k}\lambda^* _{2jk}}{(\Delta -\Delta_j 
 +\omega_k +\bar{\omega} ) } 
\: , 
& \fl \alpha_8 = \sum_{j\notin \{1,2\}} 
 \frac{2\lambda _{2jk}\lambda^* _{j2k}}{(\Delta -\Delta_j 
 -\omega_k +\bar{\omega} ) } 
\: ,\\
\fl \alpha_9 = \sum_{j\notin \{1,2\}} 
 \frac{2\lambda _{j2}\lambda^* _{2j}}{(\Delta -\Delta_j +2 \bar{\omega} ) } 
\: , 
& \fl \alpha_{10} = \sum_{j\notin \{1,2\}}  
 \frac{2\lambda _{2j}\lambda^* _{j2}}{(\Delta -\Delta_j ) } 
\: ,\\
\fl \alpha_{11} = \sum_{j\notin \{1,2\}} \Biggl (
- \frac{\lambda _{1j}\lambda _{j1}}{(\Delta_j +\omega_k -\bar{\omega} ) } 
- \frac{\lambda _{1j}\lambda _{j1}}{(\Delta_j +\omega_k +\bar{\omega} ) } 
\Biggr )\: ,\\
\fl \alpha_{12} = \sum_{j\notin \{1,2\}} \Biggl (
 \frac{\lambda _{2j}\lambda _{j2}}{(\Delta -\Delta_j ) } 
+ \frac{\lambda _{2j}\lambda _{j2}}{(\Delta -\Delta_j +2 \bar{\omega} ) } 
\Biggr)\: ,
\end{eqnarray}
and the detunings are defined as
\begin{eqnarray}
\Delta = \omega_2 - \omega_1 - (\omega_k + \bar{\omega})\: , \qquad
\Delta_j = \omega_{j1}-\omega_k\: .
\end{eqnarray}
Here, the terms proportional to $\alpha_0$ are the usual one-photon 
Hamiltonian parts. The next addends including $\alpha_1, \alpha_{2}$ 
are two-photon transitions which will turn out to be of most interest for 
the following analysis. 
Terms with  $\alpha_3, \dots, \alpha_{10}$ may be
interpreted as Stark shifts due to the presence of the two electromagnetic 
field modes. This Hamiltonian is an extension to the one obtained 
in~\cite{cardimona} 
in that it includes more transitions such as direct one-photon transitions 
due to the vacuum and also field ordering effects. These ordering effects
are the reason why for example $\alpha_3$ and $\alpha_4$ here are not the same
other than in~\cite{cardimona}.
The addends proportional to $\alpha_{11}, \alpha_{12}$ are somewhat
generalized Stark shifts in that they involve the population operators
$\sigma_{11}$ and $\sigma_{22}$. Usually they do not appear in effective
multiphoton Hamiltonians as they may be dropped in a rotating wave
approximation if the frequency of the involved photons is large enough. 
For example, the addend with $\alpha_{11}$ 
stems from a transition $b\: |j\rangle \langle 1|$ followed by a 
transition $b\: |1\rangle \langle j|$, one of which is counter-rotating.
But here $b$ represents a low-frequency photon such that these terms may not
be dropped a priori. However as discussed later 
the numerical simulations indicate that they do not disturb the effects
described in \cite{prl}. One hint that may help to understand this is that 
from a quantum mechanical point  of view, these contributions account for 
a distribution of the system state over the various Fock states of the
low-frequency field without inducing atomic transitions. 
The trapping effect however does not rely on a specific configuration of the
low-frequency field modes such as a concentration of the states to only 
few of the Fock states.

The interpretation of the general structure of this Hamiltonian 
is straightforward. In each addend of the $\alpha_l$ ($l=0,\dots,12$), 
the number of $\lambda$ coefficients
is equal to the number of photons exchanged.
The various processes involved may also be read off from the $\lambda$
coefficients, as for example in $\alpha_1$ in equation~(\ref{alpha1}), which
describes a transition of the effective atom from the ground state
$|1\rangle$ to the excited state $|2\rangle$ together with the absorption 
of both a low-frequency and a spontaneous photon ($b a_k \sigma_{21}$ in
Hamiltonian equation~(\ref{ham2_ad})). The first addend stems from a transition
between $|1\rangle$ and an intermediate state $|j\rangle$ via a 
vacuum-induced transition ($\lambda_{j1k}$) and a low frequency field photon
absorption between $|j\rangle$ and $|2\rangle$ ($\lambda_{2j}$). The
third addend is due to a transition between $|1\rangle$ and
an intermediate state $|j\rangle$ via a low frequency field photon 
absorption ($\lambda_{j1}$) and a vacuum-induced transition between 
$|j\rangle$ and $|2\rangle$ ($\lambda_{2jk}$).

\subsubsection{Three-photon Hamiltonian}
Using the results of section~\ref{higher-order} 
in the expression for the three-photon Hamiltonian equation~(\ref{h3}) yields
after a calculation as in section~\ref{sec-two-photon} for the two-photon
Hamiltonian the following three-photon Hamiltonian:
\begin{eqnarray}
\fl H_I^{(3-photon)} = H_I^{(2-photon)} 
 + \hbar \: \sum_k \: \Bigl \{ (\beta_1 bb^\dagger + \beta_2 b^\dagger b
  +\beta_3 bb + \beta_4 b^\dagger b^\dagger)\:
  a_k^\dagger \sigma_{12}
 + \textrm{ h.c.} \Bigr \} .\label{ham3_ad}
\end{eqnarray}
Thus as expected the one- and two-photon processes of this Hamiltonian 
are identical to the ones in the two-photon Hamiltonian equation~(\ref{ham2_ad}).
All additional terms in equation~(\ref{ham3_ad}) are three-photon processes. 
The terms with $\beta_1, \beta_2$ are corrections to $\alpha_0$ in the 
two-photon Hamiltonian, as they effectively induce the same transition.  
$\beta_3$ and $\beta_4$ are coefficients
to the three-photon transition parts, which will be of most interest in the
simulation of the decay dynamics. Third-order Stark effects
vanish due to symmetry reasons.
Introducing the detuning $\Delta_n = \omega_{n1}-\omega_k $,
the explicit expressions for the $\beta$ coefficients
are given as follows (Note that
the sums over $j$ and $n$ have been omitted for notational simplicity;
all occurrences of these indices are summed over all auxiliary states
in these coefficients):
\begin{eqnarray}
\fl \beta_1 = 
\frac{\lambda _{2j}\lambda^* _{12k}\lambda^* _{j2}}
	{\Delta_j (\Delta -\Delta_j +\bar{\omega} ) } 
+ \frac{\lambda _{12}\lambda^* _{2jk}\lambda^* _{j2}}
	{\Delta_j (\Delta -\Delta_j +\omega_k +2 \bar{\omega} ) } 
+ \frac{\lambda _{1j}\lambda^* _{jnk}\lambda^* _{n2}}
	{\Delta_n (\Delta_j +\omega_k ) } 
\nonumber \\ 
\fl \quad + \frac{\lambda _{2j}\lambda^* _{12}\lambda^* _{j2k}}
	{(\Delta_j +\omega_k -\bar{\omega} ) 
	 (\Delta -\Delta_j +\bar{\omega} ) } 
+ \frac{\lambda _{jn}\lambda^* _{j2}\lambda^* _{n1k}}
	{(\Delta -\Delta_n ) (\Delta -\Delta_j -\bar{\omega} ) } 
+ \frac{\lambda _{n2}\lambda^* _{1jk}\lambda^* _{jn}}
	{(\Delta_n -2 \bar{\omega} ) (\Delta_j -\bar{\omega} ) }
\nonumber \\ 
\fl\quad  + \frac{\lambda _{21}\lambda^* _{j1k}\lambda^* _{j1}}
	{(\Delta -\Delta_j ) (\Delta_j +\omega_k +2 \bar{\omega} ) } 
+ \frac{\lambda _{1j}\lambda^* _{jn}\lambda^* _{n2k}}
	{(\Delta_j +\omega_k ) (\Delta_n +\omega_k -\bar{\omega} ) } 
\nonumber \\ 
\fl\quad  + \frac{\lambda _{j1}\lambda^* _{21}\lambda^* _{j1k}}
	{(\Delta -\Delta_j -\omega_k -\bar{\omega} ) 
	 (\Delta_j +2 \omega_k +\bar{\omega} ) } 
+ \frac{\lambda _{j1}\lambda^* _{21k}\lambda^* _{j1}}
	{(\Delta -\Delta_j -\omega_k -\bar{\omega} ) 
	 (\Delta_j +\omega_k +2 \bar{\omega} ) } 
\nonumber \\ 
\fl\quad  + \frac{\lambda _{n1}\lambda^* _{j2}\lambda^* _{jnk}}
	{(\Delta -\Delta_j -\bar{\omega} ) 
	 (\Delta -\Delta_n -\omega_k -\bar{\omega} ) } %
+ \frac{\lambda _{jn}\lambda^* _{j2k}\lambda^* _{n1}}
	{(\Delta -\Delta_j -\omega_k )
	 (\Delta -\Delta_n -\omega_k +\bar{\omega} ) } 
\: ,  \\ 
\fl \beta_2 =
\frac{\lambda _{j1}\lambda^* _{21}\lambda^* _{j1k}}
	{(\Delta -\Delta_j ) (\Delta_j +\omega_k ) } 
+ \frac{\lambda _{n2}\lambda^* _{1j}\lambda^* _{jnk}}
	{(\Delta_n -2 \bar{\omega} ) (\Delta_j +\omega_k -2 \bar{\omega} ) } 
+ \frac{\lambda _{jn}\lambda^* _{1jk}\lambda^* _{n2}}
	{\Delta_n (\Delta_j -\bar{\omega} ) } 
 \nonumber \\ 
\fl\quad + \frac{\lambda _{n1}\lambda^* _{j2k}\lambda^* _{jn}}
	{(\Delta -\Delta_j -\omega_k )
	 (\Delta -\Delta_n -\omega_k -\bar{\omega} ) } 
+ \frac{\lambda _{j2}\lambda^* _{12k}\lambda^* _{2j}}
	{(\Delta_j -2 \bar{\omega} ) (\Delta -\Delta_j +3 \bar{\omega} ) }
\nonumber \\ 
\fl\quad+ \frac{\lambda _{j2}\lambda^* _{jn}\lambda^* _{n1k}}
	{(\Delta -\Delta_n ) (\Delta -\Delta_j +\bar{\omega} ) }
+ \frac{\lambda _{jn}\lambda^* _{1j}\lambda^* _{n2k}}
	{(\Delta_j +\omega_k -2 \bar{\omega} )
	 (\Delta_n +\omega_k -\bar{\omega} ) } 
\nonumber \\ 
\fl\quad
+ \frac{\lambda _{j1}\lambda^* _{21k}\lambda^* _{j1}}
	{(\Delta_j +\omega_k )
	 (\Delta -\Delta_j -\omega_k +\bar{\omega} ) } 
+ \frac{\lambda _{j2}\lambda^* _{jnk}\lambda^* _{n1}}
	{(\Delta -\Delta_j +\bar{\omega} )
	 (\Delta -\Delta_n -\omega_k +\bar{\omega} ) } 
\nonumber \\ 
\fl\quad
+ \frac{\lambda _{21}\lambda^* _{j1k}\lambda^* _{j1}}
 	{(\Delta -\Delta_j -\omega_k +\bar{\omega} )
	 (\Delta_j +2 \omega_k +\bar{\omega} ) } 
+ \frac{\lambda _{j2}\lambda^* _{12}\lambda^* _{2jk}}
	{(\Delta_j -2 \bar{\omega} )
	 (\Delta -\Delta_j +\omega_k +2 \bar{\omega} ) } 
\nonumber \\ 
\fl\quad
+ \frac{\lambda _{12}\lambda^* _{2j}\lambda^* _{j2k}}
	{(\Delta_j +\omega_k -\bar{\omega} )
	 (\Delta -\Delta_j +3 \bar{\omega} ) } 
\: ,   \\ 
\fl \beta_3 =
\frac{\lambda _{jn}\lambda _{n2}\lambda^* _{1jk}}
	{(\Delta_j -3 \bar{\omega} ) (\Delta_n -2 \bar{\omega} ) } 
+ \frac{\lambda _{1j}\lambda _{n2}\lambda^* _{jnk}}
	{(\Delta_n -2 \bar{\omega} ) (\Delta_j +\omega_k -2 \bar{\omega} ) } 
+ \frac{\lambda _{j2}\lambda _{jn}\lambda^* _{n1k}}
	{(\Delta -\Delta_n ) (\Delta -\Delta_j -\bar{\omega} ) } 
\nonumber \\ 
\fl\quad
+ \frac{\lambda _{21}\lambda _{j1}\lambda^* _{j1k}}
	{(\Delta -\Delta_j ) (\Delta_j +\omega_k +2 \bar{\omega} ) } 
+ \frac{\lambda _{2j}\lambda _{j2}\lambda^* _{12k}}
	{(\Delta_j -2 \bar{\omega} ) (\Delta -\Delta_j +3 \bar{\omega} ) } 
+ \frac{\lambda _{12}\lambda _{2j}\lambda^* _{j2k}}
	{(\Delta_j +\omega_k -\bar{\omega} )
	 (\Delta -\Delta_j +3 \bar{\omega} ) } 
\nonumber \\ 
\fl \quad
+ \frac{\lambda _{jn}\lambda _{n1}\lambda^* _{j2k}}
	{(\Delta -\Delta_j -\omega_k -2 \bar{\omega} ) (\Delta -\Delta_n -\omega_k -\bar{\omega} ) } 
+ \frac{\lambda _{j2}\lambda _{n1}\lambda^* _{jnk}}
	{(\Delta -\Delta_j -\bar{\omega} ) (\Delta -\Delta_n -\omega_k -\bar{\omega} ) } 
\nonumber \\ 
\fl \quad
+ \frac{\lambda _{1j}\lambda _{jn}\lambda^* _{n2k}}
	{(\Delta_j +\omega_k -2 \bar{\omega} )
	 (\Delta_n +\omega_k -\bar{\omega} ) } 
+ \frac{\lambda^* _{21k} \lambda_{j1} \lambda _{j1}}
	{(\Delta -\Delta_j -\omega_k -\bar{\omega} )
	 (\Delta_j +\omega_k +2 \bar{\omega} ) }
\nonumber \\* 
\fl \quad
+ \frac{\lambda _{21}\lambda _{j1}\lambda^* _{j1k}}
	{(\Delta -\Delta_j -\omega_k -\bar{\omega} )
	 (\Delta_j +2 \omega_k +3 \bar{\omega} ) } 
+ \frac{\lambda _{12}\lambda _{j2}\lambda^* _{2jk}}
	{(\Delta_j -2 \bar{\omega} )
	 (\Delta -\Delta_j +\omega_k +4 \bar{\omega} ) }  
\: ,   \label{beta3} \\ 
\fl \beta_4 =
\frac{\lambda^* _{12}\lambda^* _{2jk}\lambda^* _{j2}}
	{\Delta_j (\Delta -\Delta_j +\omega_k ) } 
+ \frac{\lambda^* _{12}\lambda^* _{2j}\lambda^* _{j2k}}
	{(\Delta_j +\omega_k -\bar{\omega} )
	 (\Delta -\Delta_j +\bar{\omega} ) } 
+ \frac{\lambda^* _{21}\lambda^* _{j1k}\lambda^* _{j1}}
	{(\Delta -\Delta_j ) (\Delta_j +\omega_k ) }
\nonumber \\ 
\fl \quad
+ \frac{\lambda^* _{1j}\lambda^* _{jn}\lambda^* _{n2k}}
	{(\Delta_j +\omega_k ) (\Delta_n +\omega_k -\bar{\omega} ) }
+ \frac{\lambda^* _{12k}\lambda^* _{2j}\lambda^* _{j2}}
	{\Delta_j (\Delta -\Delta_j +\bar{\omega} ) } 
+ \frac{\lambda^* _{j2}\lambda^* _{jn}\lambda^* _{n1k}}
	{(\Delta -\Delta_n ) (\Delta -\Delta_j +\bar{\omega} ) }  
\nonumber \\ 
\fl \quad
+ \frac{\lambda^* _{1jk}\lambda^* _{jn}\lambda^* _{n2}}
	{\Delta_n (\Delta_j +\bar{\omega} ) } 
+ \frac{\lambda^* _{21k} \lambda^*_{j1} \lambda^*_{j1}}
	{(\Delta_j +\omega_k ) (\Delta -\Delta_j -\omega_k +\bar{\omega} ) } 
+ \frac{\lambda^* _{1j}\lambda^* _{jnk}\lambda^* _{n2}}
	{\Delta_n (\Delta_j +\omega_k ) } 
\nonumber \\ 
\fl \quad
+ \frac{\lambda^* _{21}\lambda^* _{j1k}\lambda^* _{j1}}
	{(\Delta_j +2 \omega_k -\bar{\omega} )
	 (\Delta -\Delta_j -\omega_k +\bar{\omega} ) } 
+ \frac{\lambda^* _{j2}\lambda^* _{jnk}\lambda^* _{n1}}
	{(\Delta -\Delta_j +\bar{\omega} )
	 (\Delta -\Delta_n -\omega_k +\bar{\omega} ) } 
\nonumber \\ 
\fl \quad
+ \frac{\lambda^* _{j2k}\lambda^* _{jn}\lambda^* _{n1}}
	{(\Delta -\Delta_n -\omega_k +\bar{\omega} )
	 (\Delta -\Delta_j -\omega_k +2 \bar{\omega} ) } 
.
\end{eqnarray}
\begin{figure}[t]
\center
\includegraphics[height=5cm]{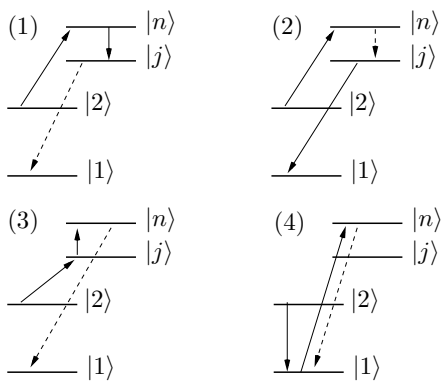}
\caption{\label{transition} The first four transitions involved
in the coupling constant $\beta_3$ corresponding to the
transition $b b a_k^\dagger \sigma_{12}$. The solid lines denote
interactions with the low-frequency field, the dashed line
denotes a vacuum photon emission. All paths start in state $|2\rangle$ and
end in state $|1\rangle$. The lengths of the arrows are not related
to the frequencies of the involved photons.}
\end{figure}
The structure of the coefficients is similar to the one in the two-photon
Hamiltonian, and the contributing transition pathways may be read off
each addend. Figure~\ref{transition} shows the first four processes 
contributing to $\beta_3$. This coefficient corresponds to the
transition $b b a_k^\dagger \sigma_{12}$ in the Hamiltonian 
equation~(\ref{ham3_ad}). 
For example, the first addend in $\beta_3$ in equation~(\ref{beta3}) describes 
a path starting by a transition from the excited state $|2\rangle$
to the auxiliary state $|n\rangle$ induced by a low-frequency field absorption,
followed by a transition to the auxiliary state $|j\rangle$
induced by a low-frequency field absorption, and then a vacuum photon
emission with a transition to the ground state $|1\rangle$. The other
transition pathways may be read off the addends accordingly.
It is important to note that the frequency of the spontaneous
photon emitted or absorbed in the multiphoton processes 
is always close to the atomic transition frequency,
because the sum of the frequencies of all photons
(with a relative sign between emitted and absorbed photons)
is required to be close to the atomic transition frequency
for a spontaneous emission to occur. This will be shown 
using a Wigner-Weisskopf-like calculation in section~\ref{decay}.

For any application of the above Hamiltonian equation~(\ref{ham3_ad}),
it is important to be able to decide which intermediate states
need to be taken into account in the analysis.
First of all, the number of exchanged photons restricts the atomic level space.
For example in a three-photon transition from a $S$ (angular momentum
$l=0$) to a $P$ ($l=1$) state, states with $l>2$ are not possible
as intermediate states.
The two important parameters which decide about the relevance 
of the remaining intermediate states are the transition 
dipole moments which connect the state in question to other
relevant states, and the frequency separation to the other states.
A reasonable parameter involving
these quantities is e.g. the ratio $\beta_i/\alpha_0$ ($i=1,\dots,4$),
i.e., the coupling strength of the multiphoton transitions relative
to the single photon transition strength.
By evaluating the expression for $\beta_i$ for each combination of
the intermediate states $j$ and $n$ separately, it is possible to 
compare the relative  weights of the various pathways as e.g. shown 
in figure~\ref{transition}.
Then the states which give rise to pathways with small relative contribution can be
neglected. For this, it is not necessary to evaluate all possible 
combinations. If $m$ is the principal quantum number of an
intermediate state which is higher than the principal quantum
numbers of the ground and the excited state,
then the contribution of the corresponding intermediate state with principal
quantum number $m+1$ can be expected to be lower 
than the contribution of the state with quantum number $m$ due to the larger 
frequency separation and due to a lower transition dipole moment.

\section{\label{decay}Decay dynamics of the effective two-level system}
\subsection{General considerations}
To further understand the various contributions in the derived Hamiltonian,
and to analyze the modification of the decay dynamics due to the additional 
multiphoton transition pathways we apply the three-photon Hamiltonian 
equation~(\ref{ham3_ad}) in a quantum mechanical simulation of the effective
two-level atom subject to an intense low-frequency field and the vacuum modes.
For this, we rearrange the free and the interaction part of the three-photon
Hamiltonian as follows:
\begin{eqnarray}
\fl H_0^{(3-photon)} = \hbar \bar{\omega} b^\dagger b 
  + \hbar \sum_k  \omega_k a_k^\dagger a_k \nonumber \\
\lo+ \hbar \: \Bigl \{ \omega_1 + \alpha_5\:b^\dagger \: b 
  + \alpha_6\: b \: b^\dagger 
  + \sum_k (\alpha_3\:a_k^\dagger \: a_k 
  + \alpha_4\:a_k \: a_k^\dagger) 
   \Bigr \}\: \sigma_{11} \nonumber  \\
\lo  + \hbar \: \Bigl \{ \omega_2 
  + \alpha_9\:b^\dagger \: b + \alpha_{10}\: b \: b^\dagger 
    + \sum_k (\alpha_{7}\:a_k^\dagger \: a_k
  + \alpha_{8}\:a_k \: a_k^\dagger) \Bigr \}\: \sigma_{22} \: ,
  \label{h0-simul} \\
\fl H_{vac}^{(3-photon)} = \hbar \: \sum_k \: \Bigl \{ \alpha_0 + \alpha_1\:b
  + \alpha_2\:b^\dagger + \beta_1^*\:b\:b^\dagger \nonumber \\
  + \beta_2^*\:b^\dagger \:b + \beta_3^*\:b^\dagger \:b^\dagger
 + \beta_4^*\:b\:b \Bigr \}\: \:a_k\:\sigma_{21} + \textrm{ h.c.}\: ,\\
\fl  H_{b}^{(3-photon)} = \hbar \: \Bigl ( \alpha_{11}\: b\:b  \:\sigma_{11} 
+ \alpha_{12}\: b\:b \:\sigma_{22} \Bigr )  + \textrm{ h.c.} \: .
\end{eqnarray}
$H_0^{(3-photon)}$ is the effective free Hamiltonian, including the Stark
shifts of the two atomic levels.
$H_{vac}^{(3-photon)}$ describes the interaction with the vacuum field, 
i.e. the spontaneous decay. 
The contributions involving operators $b$ and $b^\dagger$ in this part 
account for multiphoton transitions
consisting of one interaction with the vacuum field and one or more
interactions with the low-frequency field.
$H_{b}^{(3-photon)}$ contains terms due to the low frequency field alone
 which drive the effective system.

In order to calculate the system evolution we notice that 
$H_{b}^{(3-photon)}$ contains  multiphoton processes only such that its 
coupling constants $\alpha_i$ ($i\in \{11,12\}$) are moderate even for 
an intense low-frequency field.
Thus in the regime where the previous approximations such as the 
adiabatic elimination
are valid, the evolution which gives rise to the damping of the system 
(i.e. the coupling to the vacuum) may be evaluated 
separately from external driving fields as it is common practice 
in quantum optical calculations, see e.g. Chapter 8.6.1 in  \cite{puri}.
Thus in our case, we evaluate the system
with $H_0^{(3-photon)} + H_{vac}^{(3-photon)}$ alone and then 
combine the resulting equations of motion with the ones resulting from
 the driving part $H_{b}^{(3-photon)}$. This
approximation holds if $|| H_{b}^{(3-photon)} || \ll ||H_0^{(3-photon)} ||$, 
which means that coupling constants in $H_{b}^{(3-photon)}$ have to be low 
as compared to the atomic transition frequency
of the effective two-level system. 

First, we transfer the vacuum interaction part of the Hamiltonian
$H_{vac}^{(3-photon)}$ to the interaction picture with
respect to the free part $H_0^{(3-photon)}$. To understand the Stark shift
contribution, we transfer the partial
Hamiltonian $a_k^\dagger \: b^n \: \sigma_{21}$ as an example:
\begin{equation}
a_k^\dagger \: b^n \: \sigma_{21} \Rightarrow a_k^\dagger \: b^n \: \sigma_{21}
\:e^{-\im t(\omega_2 - \omega_1 -\omega_k + n\bar{\omega})} \: S_a \: S_b 
 \label{example-Ham}
\end{equation}
with
\begin{eqnarray}
S_a &=&  e^{-\im t\sum_l(\alpha_7 + \alpha_{8})(a_l^\dagger a_l - \delta_{kl})}
\: e^{\im t\sum_l(\alpha_3 + \alpha_{4})\: a_l^\dagger a_l} \: ,\\
S_b &=&  e^{-\im t(\alpha_9 + \alpha_{10})(b^\dagger b + n)}\: 
e^{\im t(\alpha_5 + \alpha_{6})\:b^\dagger b} 
\end{eqnarray}
as the Stark shift contributions due to spontaneous photons $S_a$ and due to 
coherent low-frequency photons $S_b$.
The two exponential functions in each of the contributions describe the 
Stark shift of the upper 
level $|2\rangle$ and of the lower level
$|1\rangle$, respectively. As expected, the shifts are proportional to 
the number of photons in the respective
modes. The difference in the photon number factors for the two states 
(e.g. $(b^\dagger b + n)$ as compared to $b^\dagger b$) is due to the
fact that the example Hamiltonian part induces changes in the photon numbers. 
To evaluate their importance, these contributions have to be compared to 
the exponential factor in equation~(\ref{example-Ham}).
There is at most one photon in the vacuum modes 
$(a_l^\dagger a_l \in \{0,1\})$, 
thus $S_a$ may be safely neglected. For the low-frequency field, it
is reasonable to assume a coherent state with a large mean number of 
photons $N$, as this represents an intense
laser field. As the relative photon number distribution width decreases as 
$N^{-1/2}$ for coherent states,
and as $0\leq n \leq 2$ in the vacuum part of the three-photon Hamiltonian,
we approximate $b^\dagger b \approx b^\dagger b + n \approx N$ and thus have
\begin{equation}
S_a S_b \approx e^{-\im t(\alpha_9 + \alpha_{10}-\alpha_5 - \alpha_{6})\:N}
\end{equation}
as the Stark shift contribution. 
It is important to note that the simulated model system does not loose its 
quantum character in adopting these approximations, which are less 
inspired by physical than by numerical reasoning, as we keep the photon
operators and the distinguishable Fock states for the electromagnetical fields.
The above argument holds for all terms in $H_{vac}^{(3-photon)}$
with the same result, thus these shifts may 
be taken care of by introducing an effective atomic transition frequency 
\begin{equation}
\omega = \omega_2 - \omega_1 + N\:(\alpha_9 + \alpha_{10}-\alpha_5 
- \alpha_{6}) \: .
\end{equation}
Then the vacuum interaction part becomes
\begin{eqnarray}
\fl V_{vac}^{(3-photon)} = \hbar \: \sum_k \: \Bigl ( \alpha_0 
+ \alpha_1\:b\:e^{-\im \bar{\omega}t} + \alpha_2\:b^\dagger\:
e^{\im \bar{\omega}t}  
  + \beta_1^*\:b\:b^\dagger
 + \beta_2^*\:b^\dagger \:b 
\nonumber \\
\lo+ \beta_3^*\:b^\dagger \:b^\dagger \:e^{2 \im \bar{\omega}t} 
+ \beta_4^*\:b\:b\:e^{-2 \im \bar{\omega}t} \Bigr )  \:
  a_k\:\sigma_{21} \:e^{\im (\omega - \omega_k) t} 
  + \textrm{ h.c.} \, .
\end{eqnarray}
This expression is equivalent to the interaction Hamiltonian in 
\cite{prl}, while we do not introduce the generalized ladder operators
$\sigma_+^{(j)}$ and $\sigma_-^{(j)}$ ($j \in \mathbb{Z}$)
here which were used in the semiclassical approximation in \cite{prl}.
We proceed with the ansatz for the wavefunction
\begin{equation}
|\Psi (t) \rangle = \sum_n \: E_n(t) \: |2,n,0\rangle 
+ \sum_n \sum_{k'}\: G_n^{k'}(t) \: |1,n,k'\rangle\: . \label{ansatz}
\end{equation}
Here the first index in the kets denotes the atomic state, the second 
slot represents the number of photons in
the low-frequency field, and the last entry is either $0$ for the 
vacuum without photons or $k$ for
a single photon in mode $k$. As described earlier,
we first derive an equation of motion for the state amplitudes due to 
the vacuum part
of the Hamiltonian $V_{vac}^{(3-photon)}$, which we denote by $E^{vac}_n$ 
and $G_n^{k, vac}$.
In a second step, we calculate the equations corresponding to
the driving part of the Hamiltonian $V_{b}^{(3-photon)}$ with 
amplitudes $E^{b}_n$.
These two sets of equations of motion are then summed to give the equations 
for the full amplitudes $E_n$ and $G_n^k$.
Inserting the ansatz equation~(\ref{ansatz}) in the Schr\"odinger equation 
with Hamiltonian $V_{vac}^{(3-photon)}$
yields as vacuum part of the equations of motion for the state amplitudes
\begin{eqnarray}
\fl \im \hbar \: \frac{d}{dt} E^{vac}_n = \langle 2,n,0|V_{vac}^{(3-photon)}
 |\Psi \rangle
 = \hbar \: \sum_k \: \Bigl \{ \left ( \alpha_0 + \beta_1^*(n+1) 
 + \beta_2^* n \right ) \: G_n^{k, vac}
\nonumber \\
\lo+ \alpha_1\:\sqrt{n+1}\: G_{n+1}^{k, vac}  \: e^{-\im \bar{\omega} t}  
 + \alpha_2 \: \sqrt{n} \: G_{n-1}^{k, vac} \: e^{\im \bar{\omega}t}  
\nonumber \\
\lo+ \beta_4^*\: \sqrt{(n+1)(n+2)}\:  G_{n+2}^{k, vac}  \: 
 e^{-2\im \bar{\omega} t}
\nonumber \\
\lo+ \beta_3^*\: \sqrt{n(n-1)}\:  G_{n-2}^{k, vac}  \: 
e^{2\im \bar{\omega} t} \Bigr \}\:
e^{\im (\omega - \omega_k) t}\:, \label{eom-en}\\
\fl \im \hbar \: \frac{d}{dt} G_n^{k, vac} = \langle 1,n,k|V_{vac}^{(3-photon)}
|\Psi \rangle 
\nonumber \\
\lo= \hbar \:  \Bigl \{ (\alpha_0^*  + \beta_1(n+1) + \beta_2 n) \: 
E_n^{vac} 
+ \alpha_1^* \: \sqrt{n} \: 
 E_{n-1}^{vac} \: e^{\im \bar{\omega}t} 
\nonumber \\ 
\lo+ \alpha_2^* \: \sqrt{n+1} \:
 E_{n+1}^{vac} \: e^{-\im \bar{\omega}t}
 + \beta_4\: \sqrt{n(n-1)}\: E_{n-2}^{vac}\:e^{2\im \bar{\omega}t} 
 \nonumber \\
\lo+ \beta_3\: \sqrt{(n+1)(n+2)}\: E_{n+2}^{vac}\:e^{-2\im \bar{\omega}t}
   \Bigr \}\:e^{-\im (\omega - \omega_k) t}\: .   \label{eom-gkn}
\end{eqnarray}
Formally integrating (\ref{eom-gkn}) and inserting the result in 
(\ref{eom-en}) allows to obtain an equation of motion for the upper state
amplitudes $E_n^{vac}$ of the form
\begin{equation}
\frac{d}{dt} E_n^{vac} = - \int_0^t \sum_k \sum_{l=n-4}^{n+4} 
C_{l}(t,t')  E_{l}^{vac}(t') \: dt' \label{wigner}
\end{equation}
where $C_{l}(t,t')$ are time dependent coefficients depending on the 
coupling constants $\alpha_i$ and $\beta_j$. 
Expanding equation~(\ref{wigner}), it turns out that all addends on the right 
hand side are of the form
\begin{eqnarray}
\fl A = - \int_0^t \sum_k  R(\omega_k)\lambda_{rsk}\:S^*(\omega_k) 
\lambda^*_{xyk} e^{\im \mu \bar{\omega} t} e^{\im \nu \bar{\omega} t'}
  e^{\im (\omega - \omega_k + \kappa \bar{\omega})(t-t')}
E_{n+\delta}^{vac}(t') \: dt' \: .
\end{eqnarray}
Here, $r,s,x,y,\mu,\nu,\kappa$ and $\delta$ are integers depending on 
the specific addend. $R(\omega_k)\lambda_{rsk}$ and 
$S(\omega_k) \lambda_{xyk}$ are one of the coefficients 
$\alpha_i$ ($i=0,1,2$) or $\beta_j$ ($j=1,2,3,4$) respectively, written 
as products of the dipole moment corresponding to the spontaneous transition
$\lambda_{rsk}$ and $\lambda_{xyk}$ and 
of the remaining factors $R(\omega_k)$ and $S(\omega_k)$.
In a Wigner-Weisskopf-like calculation, this generic contribution may 
be evaluated to give
\begin{eqnarray}
\fl A = -  \frac{R(\omega + \kappa \bar{\omega})S^*(\omega + \kappa
\bar{\omega})\: \vec{d}_{rs}\vec{d}_{xy}}{6\hbar \epsilon_0 c^3 \pi}
(\omega+\kappa \bar{\omega})^3 \: 
    E_{n+\delta}^{vac}(t) \: e^{\im (\mu + \nu) \bar{\omega} t}
 \nonumber \\
\fl\quad = -  R(\omega + \kappa \bar{\omega})S^*(\omega + \kappa \bar{\omega}) \:
 p^{xy}_{rs}
\: e^{\im (\mu + \nu) \bar{\omega} t} 
\: \frac{1}{2} \: \sqrt{\Gamma_{rs}(\omega + \kappa
 \bar{\omega})\Gamma_{xy}(\omega + \kappa \bar{\omega})}\:
  E_{n+\delta}^{vac}(t) \: .
\label{specific}
\end{eqnarray}
Here $\Gamma_{ij}(x)$ is the spontaneous decay rate of transition
 $i\leftrightarrow j$ with
transition frequency modified to $x$, i.e. 
\[
\Gamma_{ij}(x) = \left ( \frac{1}{4\pi \epsilon_0} 
\frac{4 |d_{ij}|^2 \omega_{ij}^3}{3\hbar c^3} \right ) \:
 \frac{x^3}{\omega_{ij}^3}
= \hat{\Gamma}_{ij}\frac{x^3}{\omega_{ij}^3}\, , 
\]
where $\hat{\Gamma}_{ij}$ is the spontaneous decay rate of transition
 $i\leftrightarrow j$ 
with transition frequency $\omega_{ij}$, and
\begin{equation}
p^{xy}_{rs} = \frac{\vec{d}_{rs}\vec{d}_{xy}}{|\vec{d}_{rs}|\cdot
 |\vec{d}_{xy}|}
\end{equation}
is a prefactor describing the amount of quantum interference possible 
between transition $r\leftrightarrow s$ and
transition $x\leftrightarrow y$. It is zero if the dipole moments are
orthogonal and reaches its maximum value of 1
(-1) for parallel (antiparallel) dipole moments. Thus we find the usual form 
of a square root of the product of the two
corresponding spontaneous decay rates as characteristic for vacuum induced
interference effects \cite{int2b}.
The $R$ and $S$ merely are prefactors which are present because the
corresponding process is a multiphoton process.
Assuming as for the Stark shift contributions that the photon number
distribution width of the 
low-frequency field is negligible as compared to the number of photons $N$, 
i.e. $n \approx n+1 \approx n-1 \approx \dots \approx N$,
one may introduce the low-frequency field Rabi frequency in equation~(\ref{wigner})
given by
\begin{equation}
\Omega_{ij} = 2\: \lambda_{ij}\sqrt{n+1}
\end{equation}
for transition $i\leftrightarrow j$.
This finally leads to a system of equations for the upper state 
populations given by
\begin{eqnarray}
\fl \frac{d}{dt} E_n^{vac}(t)  =  -c_0 \: E_n(t) 
- c_{1} \: e^{\im \bar{\omega} t}\: E_{n-1}^{vac}(t) 
- c_{2} \: e^{-\im \bar{\omega} t}\: E_{n+1}^{vac}(t) 
- c_{3} \: e^{2 \im \bar{\omega} t}\: E_{n-2}^{vac}(t) 
\nonumber \\
\lo- c_{4} \: e^{-2 \im \bar{\omega} t}\: E_{n+2}^{vac}(t) 
- c_{5} \: e^{3 \im \bar{\omega} t}\: E_{n-3}^{vac}(t)
- c_{6} \: e^{-3 \im \bar{\omega} t}\: E_{n+3}^{vac}(t) 
\nonumber \\
\lo- c_{7} \: e^{4 \im \bar{\omega} t}\: E_{n-4}^{vac}(t)  
 - c_{8} \: e^{-4 \im \bar{\omega} t}\: E_{n+4}^{vac}(t) \: , \label{vac-part}
\end{eqnarray}
where the $c_i$ ($i=1\dots 8$) are constant coefficients, whose specific 
form which may be obtained by expanding equation~(\ref{wigner}) and 
(\ref{specific}) is omitted here
as e.g. $c_0$ contains of several hundred addends in its most general form. 
The generalized Stark shift contributions in the driving part
$V_{b}^{(3-photon)}$ of the Hamiltonian
give rise to an equation of motion for the upper atomic state amplitude 
which is given by
\begin{eqnarray}
\fl \im \frac{d}{dt} E^b_n(t) = \langle 2,n,0|V_{b}^{(3-photon)}(t)|\Psi(t) 
\rangle 
 = \: d_3\: e^{2 \im \bar{\omega} t}\: E_{n-2}^b(t) + d_4\: 
 e^{-2 \im \bar{\omega} t}\: E_{n+2}^b(t) \label{driving-part}
\end{eqnarray}
with constant coefficients $d_i$ ($i=3,4$).
Thus the total equation of motion  of the upper atomic state amplitude  
$E_n$ is given by the sum of the right hand side of the two contributions of
equations~(\ref{vac-part}) and (\ref{driving-part}) upon replacement of the
amplitudes $E_n^{vac}$ and $E_n^{b}$ by $E_n$:
\begin{eqnarray}
\fl\frac{d}{dt} E_n(t) = -c_0 \: E_n(t) 
 - c_{1}  \: e^{\im \bar{\omega} t}\: E_{n-1}(t) 
 - c_{2} \: e^{-\im \bar{\omega} t}\: E_{n+1}(t)  
 - (c_{3}  + \im d_3) \: e^{2 \im \bar{\omega} t}\: 
  E_{n-2}(t) \nonumber \\
\lo- (c_{4} + \im d_4) \: e^{-2 \im \bar{\omega} t}\: E_{n+2}(t) 
 - c_{5}  \: e^{3 \im \bar{\omega} t}\: E_{n-3}(t)
 - c_{6}  \: e^{-3 \im \bar{\omega} t}\: E_{n+3}(t) \nonumber \\
\lo- c_{7} \: e^{4 \im \bar{\omega} t}\: E_{n-4}(t)  
 - c_{8} \: e^{-4 \im \bar{\omega} t}\: E_{n+4}(t) \: . 
 \label{simul-eom}
\end{eqnarray}
These equations are a generalization of equations~(3) in \cite{prl}, which 
means that they have the same structure, but different coefficients $c_i$
($i=0,\dots,8$) and $d_i$ ($i=3,4$) than in \cite{prl}. The difference is 
due to the fact that 
here we include Stark shifts and do not use the semiclassical sum over all
possible low-frequency photon numbers as in \cite{prl}. Some effects of these
differences will be discussed in section~\ref{results}. In particular, 
it will turn out later that the driving terms which may be interpreted as 
generalized Stark shift terms do not change the decay dynamics
of the simulated system in a notable manner.

To simulate the system behavior, one has to choose an initial
number of photons $N$ in the low-frequency mode and a possible
range of deviations $\Delta_N$ from this number. The
set of state amplitudes considered in the analysis is then chosen as 
$\{ E_n(t) | N-\Delta_N \leq n \leq N+\Delta_N \}$. 
In the equations of motion for these state amplitudes according
to equation~(\ref{simul-eom}) all references to states outside the
chosen set are neglected. Thus $\Delta_N$ has to be chosen large enough
such that the outermost states are barely populated during the 
calculated evolution time to avoid numerical artefacts due
to the artificially added borders of the simulated level
space. 
As initial condition we choose
\begin{equation}
 |\Psi(0)\rangle = |2, \alpha, 0\rangle = \frac{1}{P}\: 
 e^{-|\alpha|^2/2}\: \sum_{n=N-W}^{N+W} \frac{\alpha^n}{\sqrt{n!}} \: 
 |2, n, 0\rangle \:. \label{initial}
\end{equation}
Thus the atom is in the excited state, the vacuum is assumed empty
and the low frequency field is in a coherent state $|\alpha\rangle$ which
simulates  a strong quantized laser field. The field parameter $\alpha$ 
is given by
\begin{equation}
\alpha = \sqrt{N}\: e^{\im \varphi} \, ,
\end{equation}
as $|\alpha|^2$ is the expectation value of the photon number in a
coherent state. The phase
$\varphi$ accounts for the possible complexity of $\alpha$.
As only upper state amplitudes $E_n(t)$ with 
$N-\Delta_N \leq n \leq N+\Delta_N$
are considered in the analysis, also the initial photon number distribution
of the coherent state has to be restricted. Thus $W$ is a cutoff of the 
photon number distribution width chosen such that $W<\Delta_N$. 
Again, this avoids population losses at the borders 
of the simulated Hilbert space. $P$ is a normalization
constant such that $\langle \Psi(0) | \Psi(0) \rangle = 1$, 
which is required because of the cutoff width $W$.
As a rough consistency check, the range parameters 
$\Delta_N$ and $W$ must be chosen large
enough for the specific system parameters such that increasing
these values does not affect the result.
After solving this set of ($2\Delta_N + 1$) coupled ordinary 
differential equations the total upper state population may be obtained
as
\begin{equation}
\Pi(t) = \sum_{n=N-\Delta_N}^{N+\Delta_N} |E_n(t)|^2\:.
\end{equation}
This population may then be compared to the exponential
decay for the two-state system without an additional
low-frequency field.

\subsection{\label{rubidium}Rubidium as an example}
\subsubsection{\label{model}Model system}

\begin{figure}[t]
\center
\includegraphics[height=5cm]{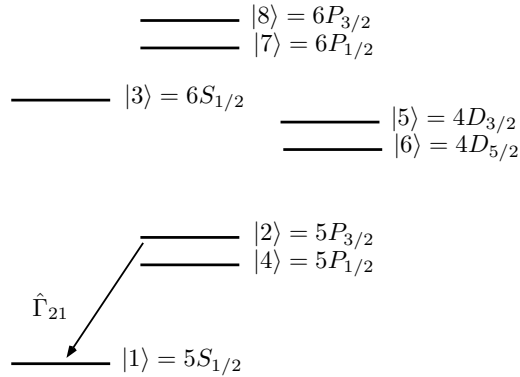}
\caption{\label{pic-rubidium} Partial level scheme of rubidium taken as 
the simulated model system.
The spontaneous emission of transition $|2\rangle \rightarrow |1\rangle$ 
is to be slowed down by the discussed scheme.
The figure is not drawn to scale.}
\end{figure}

To demonstrate the feasibility of the scheme, we use rubidium as our 
model system. The simulated atomic levels are shown in 
figure~\ref{pic-rubidium}. 
The population is assumed to be in the $5P_{3/2}$ state initially. 
Without additional fields, this state decays to the $5S_{1/2}$ state with 
a decay rate of $\hat{\Gamma}_{21} = 37.5 \cdot 10^6 \textrm{ s}^ {-1}$ 
\cite{rubidium-data}.
As discussed in the previous section, within the adiabatic approximation the
intermediate states are never populated, see equation~(\ref{aux-pop}). Thus
it is possible to include states even if there 
are decay channels leading out of the simulated level space.
Also, low spontaneous decay rates do not in general mean
that the corresponding transitions do not need to be taken into account,
as the low rates may be due to the small energy spacing between two states.
The dipole moments however which are important for the low-frequency field
may be comparable to the other transition dipole moments.
In the simulation, only dipole-allowed transitions are considered.
This amounts to a simplification of the equation of the upper 
state amplitudes equation~(\ref{simul-eom}) by e.g. eliminating 
$c_1$ and $c_2$ which rely on dipole-forbidden transitions,
as in addition to the spontaneous photon a second photon is
exchanged with the low-frequency field.
The energies of the various states are given by \cite{rubidium-data}
\begin{eqnarray*}
\mathcal{E}_1 =0 \: \textrm{eV}\: ,   \qquad & \mathcal{E}_2 =1.589 \: \textrm{eV} 
\: ,\\
\mathcal{E}_3 =2.496 \: \textrm{eV} \: , \qquad &\mathcal{E}_4 =1.560 
\: \textrm{eV}\: , \\ 
\mathcal{E}_5 =2.400 \: \textrm{eV} \: , \qquad  & \mathcal{E}_6 =2.400 
\: \textrm{eV}\: , \\
\mathcal{E}_7 =2.940 \: \textrm{eV} \: , \qquad  & \mathcal{E}_7 =2.950 
\: \textrm{eV}\: . 
\end{eqnarray*}
The energy separation $(\mathcal{E}_5 - \mathcal{E}_6)$ is about 55 MHz;
the main transition has a frequency of 
$\omega_{21} = 3.84 \cdot 10^{14}\: \textrm{Hz}$.
We consider decay rates as obtained by theoretical
calculations in \cite{rubidium-data}, where also  
the required branching ratios of the various decay 
pathways from the excited states are given:
\begin{eqnarray*}
\hat{\Gamma}_{21} =37.5 \cdot 10^6 \: \textrm{ s}^{-1}\: , 
   \qquad & \hat{\Gamma}_{41} =35.6  \cdot 10^6 \: \textrm{ s}^{-1} \: ,\\
\hat{\Gamma}_{32} =12.9  \cdot 10^6 \: \textrm{ s}^{-1}\: , 
   \qquad & \hat{\Gamma}_{34} =6.6  \cdot 10^6 \: \textrm{ s}^{-1}\: , \\
\hat{\Gamma}_{52} =2.0  \cdot 10^6 \: \textrm{ s}^{-1}\: , 
   \qquad & \hat{\Gamma}_{54} =10.7  \cdot 10^6 \: \textrm{ s}^{-1}\: , \\
\hat{\Gamma}_{62} =11.9 \cdot 10^6 \: \textrm{ s}^{-1}\: , 
   \qquad & \hat{\Gamma}_{71} =2.4  \cdot 10^6 \: \textrm{ s}^{-1}\: , \\
\hat{\Gamma}_{73} =4.3  \cdot 10^6 \: \textrm{ s}^{-1}\: , 
   \qquad & \hat{\Gamma}_{75} =2.4  \cdot 10^6 \: \textrm{ s}^{-1}\: , \\
\hat{\Gamma}_{81} =2.8  \cdot 10^6 \: \textrm{ s}^{-1}\: , 
   \qquad & \hat{\Gamma}_{83} =4.5  \cdot 10^6 \: \textrm{ s}^{-1}\: , \\
\hat{\Gamma}_{85} =0.2  \cdot 10^6 \: \textrm{ s}^{-1}\: , 
   \qquad & \hat{\Gamma}_{86} =1.7  \cdot 10^6 \: \textrm{ s}^{-1}\: .
\end{eqnarray*}
For the transitions which are listed in~\cite{rubidium-data-2},
the above values are in reasonable agreement with the data reported there.
The system parameters for the figures are chosen as
$N = 10^6$, $\Delta_N = 15000$, $W = 500$, $\phi = 0$,
$\bar{\omega} = 0.1$ MHz and $p^{xy}_{rs}=1$
if not stated otherwise.
\begin{figure}[t]
\center
\includegraphics[height=5cm]{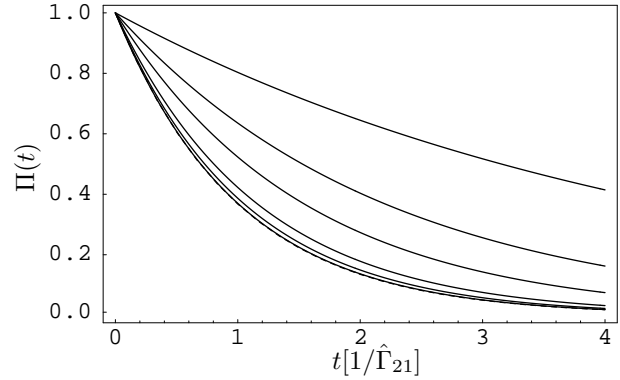}
\caption{\label{pic-om} Dependence of the total upper
state population on $\Omega_{12}$. The chosen values are
$\Omega_{12} = 1; 3; 5; 8; 10; 12 \: [10^{12}\: \textrm{Hz}]$.
The graph order corresponds to the Rabi frequency where
the highest graph corresponds to the largest driving strength.
The graph for $\Omega_{12} = 1 \cdot 10^{12}\: \textrm{Hz}$
is almost on top of the dashed reference curve.}
\end{figure}
Here $N$ only affects the initial population of the
low-frequency Fock states. It is chosen independent of
the field strength of the low-frequency field in order to
estimate the dependence of the population dynamics on the initial
conditions.
The value $p^{xy}_{rs}=1$ has been chosen for simplicity,
but is not required to for our scheme to work. 
A calculation with a $p$-value of 0.5 reduces the
trapping duration e.g. of the top curve in figure~\ref{pic-om} by less than a
factor of 3, such that the remaining effect is still considerable.
For this one should note that other than in most previously 
studied systems exhibiting quantum interference effects 
\cite{interference1,interference2,int2b,interference3},
in our setup non-zero values for $p^{xy}_{rs}$ can be found in 
any atomic system, which is due to the fact that here the two
transitions do not need to have a similar transition frequency 
or a common atomic state.
To speed up the numerical calculations, the driving terms in 
equation~(\ref{driving-part}) 
are suppressed by the replacement 
\begin{equation}
d_i \rightarrow \varrho \: d_i \:  (i=3,4) \label{damping}
\end{equation}
in equation~(\ref{simul-eom}), 
where we choose $\varrho= 1/1000$ in the following calculations.
For this one should note that for the above parameters
the simulation consists of $2\Delta_N+1 = 30001$ coupled complex
differential equations. The validity of this 
suppression is discussed in section~\ref{results}.
In the following figures~\ref{pic-om}-\ref{pic-check}, dashed lines are reference curves given by
$\exp(-\hat{\Gamma}_{21} t)$.

\begin{figure}[t]
\center
\includegraphics[height=5cm]{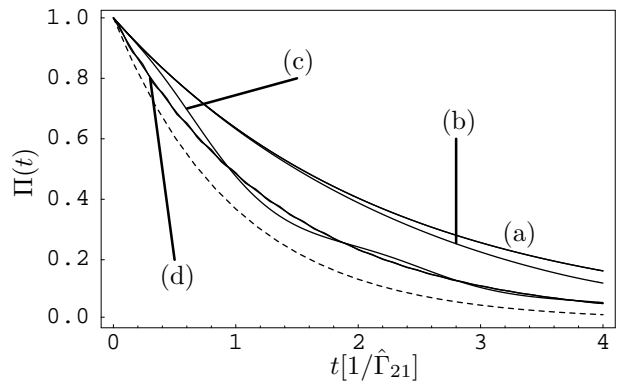}
\caption{\label{pic-frequenz} Dependence of the population trapping on
the low-frequency field frequency $\bar{\omega}$. Chosen values
are $\bar{\omega} = 10^2; 10^4; 10^6; 10^7; 10^8; 10^9 \: \textrm{Hz}$. 
(a) $\bar{\omega} = 10^2; 10^4 \: \textrm{Hz}$; 
(b) $\bar{\omega} = 10^6 \: \textrm{Hz}$ ; 
(c) $\bar{\omega} = 10^7 \: \textrm{Hz}$ and 
(d) $\bar{\omega} = 10^8;10^9 \: \textrm{Hz}$. 
The two curves in (d) can be distinguished as the one corresponding
to the lower frequency $\bar{\omega}$ slightly wiggles.
The dashed curve is the reference.}
\end{figure}

\subsubsection{\label{results}Results}
The first  figure~\ref{pic-om} shows the total upper state 
population $\Pi(t)$ for
different values of $\Omega_{12}$. 
The chosen values for this Rabi frequency are
 $\Omega_{12} = 1; 3; 5; 8; 10; 12 \: [10^{12}\: \textrm{Hz}]$.
As expected an increasing Rabi frequency of the
low-frequency field increases the amount of trapping.
The reason for this is that the relative probability
of low-frequency-field-assisted transitions increases
with an increasing field strength.
However the Rabi frequency must not be chosen too high 
as otherwise the adiabatic approximation is not valid any
longer. Also higher-order processes with more than three exchanged
photons have to be considered if the Rabi-frequency is chosen too large.
The Rabi frequency $\Omega_{12} = 10^{13} \: \textrm{Hz}$
chosen in most of the figures is about 1/40 of the atomic transition 
frequency $\omega_{21}$, which is well within the validity
range of the applied approximations.

Figure \ref{pic-frequenz} shows the role of the low-frequency field
frequency $\bar{\omega}$. The chosen values are 
$\bar{\omega} = 10^2; 10^4; 10^6; 10^7; 10^8; 10^9 \: \textrm{Hz}$, while 
the other parameters
are chosen as in figure~\ref{pic-om} with $\Omega_{12} = 10^{13}\: \textrm{Hz}$. 
As long as $\bar{\omega}$ is low as compared to the natural decay
width $\hat{\Gamma}_{12}$, the result is independent of $\bar{\omega}$
and the population plots are on top of each other. For higher 
frequencies, the population trapping decreases until the oscillation
due to $\bar{\omega}$ is visible for 
$\bar{\omega} \gtrapprox \hat{\Gamma}_{12}$.
The behavior of the system in the limit of small frequencies $\bar{\omega}$
is somewhat different from the behavior found in \cite{prl}. There the
trapping was found to improve with decreasing $\bar{\omega}$ until
in the limit $\bar{\omega} \rightarrow 0$ the decay was completely stopped.
As already discussed in \cite{prl}, this is the expected behavior
for the system simulated in \cite{prl} which is equivalent to a system with 
near-degenerate upper states \cite{zhu}. But while in our model in 
\cite{prl} all
other parameters were kept constant in changing $\bar{\omega}$, in
the present analysis other parameters such as the various
coupling strengths also change with $\bar{\omega}$, thus leading
to a different behavior. 
Still the slowing down of the spontaneous emission is most pronounced 
for low values of $\bar{\omega}$. However this is a 
key ingredient of all similar quantum interference effects \cite{zhu}
which may be depicted as follows: The smaller the field frequency 
is, the harder is it to distinguish between the various 
interference pathways, which leads to stronger quantum 
interference effects.
As the scheme relies on the fact that photons with nonzero 
frequency may be exchanged during atomic transitions, the singular
case of zero frequency, i.e. a static field, is excluded from
our analysis as in \cite{prl}.

\begin{figure}[t]
\center
\includegraphics[height=5cm]{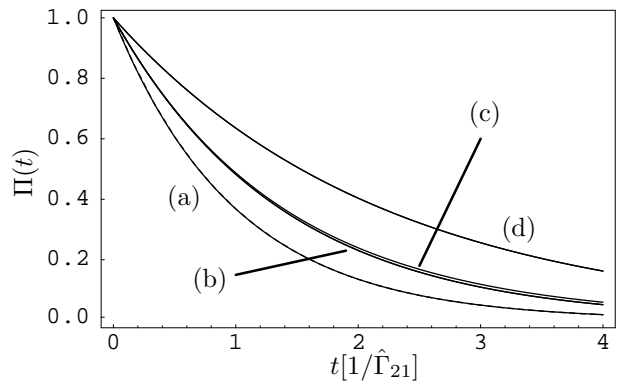}
\caption{\label{pic-phase} Dependence of the population trapping on
the phase of the initial low-frequency field coherent state.
The chosen values are $ \phi = -0.25; 0; 0.25; 0.5; 0.75; 1 [\pi]$.
Plots  are (a) for $\phi = 0.5 \pi$ and the dashed reference, (b) for 
$\phi = -0.25 \pi; 0.75 \pi$, (c) for $\phi = 0.25 \pi$, and
(d) for $\phi = 0; \pi$.}
\end{figure}

In figure~\ref{pic-phase}, the  phase $\phi$ of the initial 
low-frequency field coherent state in equation~(\ref{initial}) is varied.
The chosen values are $ \phi = -0.25; 0; 0.25; 0.5; 0.75; 1 [\pi]$; the
other parameters are chosen as in 
 figure~\ref{pic-om} with $\Omega_{12} = 10^{13}\: \textrm{Hz}$. 
The phase of the initial coherent state crucially influences the
decay dynamics of the effective two-level system.
By choosing the phase, the effective spontaneous decay varies
between the usual decay rate $\hat{\Gamma}_{21}$ and the maximum 
trapping for $\phi =  0; \pi$.
This may be understood from equation~(\ref{simul-eom}) which shows
that the decay dynamics of the state amplitude $E_n$
depends on the neighboring amplitudes $E_{n\pm m}$ ($m = 1,\dots,4$)
which change relative to $E_n$  with varying phase $\phi$.

\begin{figure}[t]
\center
\includegraphics[height=5cm]{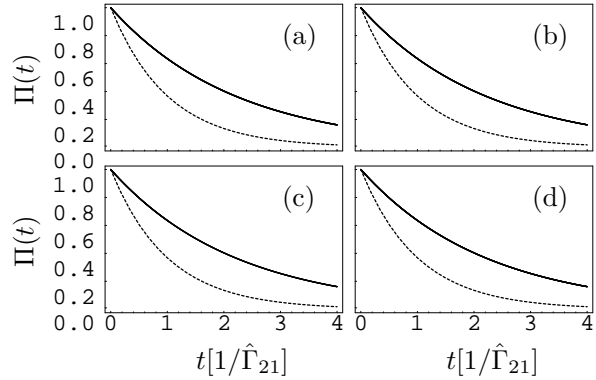}
\caption{\label{pic-check} Consistency checks of the simulation.
(a) Dependence of the population trapping on
the initial laser distribution width $W$.  The chosen values
are $W = 100, 300, 500, 1000, 2000, 3000$.
(b) Dependence of the population trapping on
the number of photons in the low-frequency mode considered in
the simulation. The chosen values
are $N = 10^4, 10^5, 10^6, 10^7, 10^8, 10^9$.
(c) Dependence of the population trapping on
the driving strength considered in
the simulation. The chosen values
are $\varrho = 10^{-5}; 10^{-4}; 10^{-3}; 10^{-2}; 10^{-1}; 1$. 
(d) Dependence of the population trapping on
the number of simulated photon number states. The chosen values
are $\Delta_N = 3000,5000,8000,10000,12000,15000$. In all subfigures,
all plots but the dashed reference are on top  of each other.
}
\end{figure}

Figure \ref{pic-check} (a) shows that the initial photon 
distribution width of the coherent low-frequency laser field
does not influence the result of the numerical calculation
notably.
This is a consistency check of the approximate coherent
state in the ansatz for the wavefunction equation~(\ref{initial}). 
The chosen values are $W = 100, 300, 500, 1000, 2000, 3000$; the
other parameters are chosen as in 
figure~\ref{pic-om} with $\Omega_{12} = 10^{13}\: \textrm{Hz}$. 


Figure \ref{pic-check} (b) shows 
that the number of photons of the coherent low-frequency laser field
does not visibly influence the result of the numerical simulation.
To evaluate the influence of the initial conditions, this value was 
chosen independent from the intensity of the low-frequency field. 
The chosen values are $N = 10^4, 10^5, 10^6, 10^7, 10^8, 10^9$; the
other parameters are chosen as in 
figure~\ref{pic-om} with $\Omega_{12} = 10^{13}\: \textrm{Hz}$. 
Increasing $N$ broadens the initial photon number distribution.
However if $N$ is large enough, neighboring photon number
states have similar amplitudes up to a possible phase.
This does not change notably for increasing $N$, and as the
equation of motion for state amplitude $E_n$ 
depends on the neighboring amplitudes $E_{n\pm m}$ ($m = 1,\dots,4$)
as discussed before, the results are independent of $N$ if it
is not too small. 


Figure~\ref{pic-check} (c) shows the dependence of the population trapping 
on the damping factor $\varrho$ of the driving terms which was introduced in 
equation~(\ref{damping}) to speed up the numerical calculations.
The chosen values are 
$\varrho = 10^{-5}; 10^{-4}; 10^{-3}; 10^{-2}; 10^{-1}; 1$; the
other parameters are chosen as in 
figure~\ref{pic-om} with $\Omega_{12} = 10^{13}\: \textrm{Hz}$. 
The Figure does not show a dependence on the damping of the
driving strength. This may be explained along the lines of the 
interpretation of figures~\ref{pic-check} (a) and (b), 
as these driving terms
merely account for a broadening of the photon number distribution 
of the low-frequency field. The trapping mechanism however does
not rely on a specific distribution width. Also, the driving terms
are due to the same interactions as the ones which lead to the 
Stark shift discussed before. 
The last consistency check is shown in figure~\ref{pic-check} (d),
where the number of simulated photon number states $\Delta_N$
is varied. The chosen values are 
$\Delta_N = 3000,5000,8000,10000,12000,15000$; the
other parameters are chosen as in 
figure~\ref{pic-om} with $\Omega_{12} = 10^{13}\: \textrm{Hz}$.
Again, there is no visible dependence on $\Delta_N$ for the chosen values which
shows that  $\Delta_N = 15000$ is large enough to eliminate
errors due to border losses.

\section{Summary}
In summing up, we have derived an explicit expression
for the multiphoton Hamiltonian describing the interaction
of an atomic two-level system with
both the vacuum field and an additional intense
low-frequency laser field including up to three-photon processes.
This Hamiltonian was used in a quantum mechanical
simulation of the decay dynamics of a two level atom
subject to the intense low-frequency laser field
and the vacuum field modes.
Using this simulation it was shown that the usual 
spontaneous decay found
on one of the transitions in rubidium may be
decelerated considerably by a suitably chosen
low-frequency field. 

\ack
Support by the German National Academic Foundation
for JE and funding by Deutsche Forschungsgemeinschaft (SFB 276) 
is gratefully acknowledged. 
We thank U.D. Jentschura
for helpful discussions related to the parameters
of the chosen model system.

\appendix
\section{\label{anhang-trans-op}First order transition operators}
In this appendix explicit expressions for the first-order 
transition operators calculated in section~\ref{higher-order} 
are given. These transition operators may be
written as
\begin{eqnarray*}
\sigma_{1j}^{(1)} &=& \sigma_{1j}^{(0)} + A_{1j}^{22} + A_{1j}^{21} 
   + \sum_{n\notin \{1,2\}} A_{1j}^{11}(n) + A_{1j}^{12}(n)\: ,\\
\sigma_{2j}^{(1)} &=& \sigma_{2j}^{(0)} + A_{2j}^{11} 
+ A_{2j}^{12} + \sum_{n\notin \{1,2\}} A_{2j}^{22}(n) + A_{2j}^{21}(n)\:.
\end{eqnarray*}
The corresponding operators $\sigma_{j1}^{(1)}$ and $\sigma_{j2}^{(1)}$ 
may be obtained by conjugation. The operators $A^{kl}_{ij}$ ($i=1,2$) 
are proportional to $\sigma_{kl}$. They are given by

\begin{equation}
\fl  A_{1j}^{11}(n) =  \mathcal A(1,n)\,,
\end{equation}
where
\begin{eqnarray}
\fl {\mathcal A}(\alpha,n) =  
 \frac{\lambda _{jnk}\lambda^* _{n\alpha k}\: a_k^\dagger\: a_k\: \sigma_{\alpha\alpha}}
 {( \omega_\alpha  -\omega_j ) ( \omega_\alpha  -\omega_n -\omega_k ) } 
+ \frac{\lambda _{n\alpha k}\lambda^* _{jnk}\: a_k\: a_k^\dagger\: \sigma_{\alpha\alpha}}
{( \omega_\alpha   -\omega_j ) ( \omega_\alpha   -\omega_n +\omega_k ) } 
\nonumber \\ 
\lo+ \frac{\lambda _{jn}\lambda^* _{n\alpha }\: b^\dagger\: b\: \sigma_{\alpha\alpha}}
 {( \omega_\alpha   -\omega_j ) ( \omega_\alpha   -\omega_n -\bar{\omega} ) } 
+ \frac{\lambda^* _{jn}\lambda^* _{n\alpha }\: b^\dagger\: b^\dagger\: \sigma_{\alpha\alpha}}
{( \omega_\alpha   -\omega_j -2 \bar{\omega} ) ( \omega_\alpha   -\omega_n -\bar{\omega} ) } \nonumber \\ 
\lo+ \frac{\lambda^* _{jn}\lambda^* _{n\alpha k}\: a_k^\dagger\: 
 b^\dagger\: \sigma_{\alpha\alpha}}
 {( \omega_\alpha   -\omega_n -\omega_k ) ( \omega_\alpha   -\omega_j 
 -\omega_k -\bar{\omega} ) } 
+ \frac{\lambda^* _{jnk}\lambda^* _{n\alpha }\: b^\dagger\: 
 a_k^\dagger\: \sigma_{\alpha\alpha}}
{( \omega_\alpha   -\omega_n -\bar{\omega} ) 
( \omega_\alpha   -\omega_j -\omega_k -\bar{\omega} ) } 
\nonumber \\ 
\lo+ \frac{\lambda _{n\alpha k}\lambda^* _{jn}\: a_k\: b^\dagger\: \sigma_{\alpha\alpha}}
 {( \omega_\alpha   -\omega_n +\omega_k ) ( \omega_\alpha   -\omega_j 
 +\omega_k -\bar{\omega} ) } 
+ \frac{\lambda _{jnk}\lambda^* _{n\alpha }\: b^\dagger\: a_k\: \sigma_{\alpha\alpha}}
{( \omega_\alpha   -\omega_n -\bar{\omega} ) ( \omega_\alpha   -\omega_j
 +\omega_k -\bar{\omega} ) } 
\nonumber \\ 
\lo+ \frac{\lambda _{n\alpha }\lambda^* _{jn}\: b\: b^\dagger\: \sigma_{\alpha\alpha}}
 {( \omega_\alpha   -\omega_j ) ( \omega_\alpha   -\omega_n +\bar{\omega} ) } 
+ \frac{\lambda _{jn}\lambda^* _{n\alpha k}\: a_k^\dagger\: b\: \sigma_{\alpha\alpha}}
{( \omega_\alpha   -\omega_n -\omega_k ) ( \omega_\alpha   -\omega_j -\omega_k
 +\bar{\omega} ) } 
\nonumber \\ 
\lo+ \frac{\lambda _{n\alpha }\lambda^* _{jnk}\: b\: a_k^\dagger\: \sigma_{\alpha\alpha}}
 {( \omega_\alpha   -\omega_n +\bar{\omega} ) ( \omega_\alpha   -\omega_j -\omega_k
  +\bar{\omega} ) } 
+ \frac{\lambda _{jn}\lambda _{n\alpha k}\: a_k\: b\: \sigma_{\alpha\alpha}}
{( \omega_\alpha   -\omega_n +\omega_k ) ( \omega_\alpha   -\omega_j +\omega_k
 +\bar{\omega} ) } 
\nonumber \\ 
\lo+ \frac{\lambda _{jnk}\lambda _{n\alpha }\: b\: a_k\: \sigma_{\alpha\alpha}}
 {( \omega_\alpha   -\omega_n +\bar{\omega} ) ( \omega_\alpha   -\omega_j 
 +\omega_k +\bar{\omega} ) } 
+ \frac{\lambda _{jn}\lambda _{n\alpha }\: b\: b\: \sigma_{\alpha\alpha}}
{( \omega_\alpha   -\omega_n +\bar{\omega} ) ( \omega_\alpha   -\omega_j 
+2 \bar{\omega} ) } \: , 
\end{eqnarray}
\begin{eqnarray}
\fl  A_{1j}^{22} = {\mathcal B}(1,2) \,,
\end{eqnarray}
where
\begin{eqnarray}
\fl {\mathcal B}(\alpha,\beta) = 
- \frac{\lambda _{\beta \alpha k}\lambda^* _{j\beta k}\: a_k\: a_k^\dagger\: \sigma_{\beta \beta }}
{( \omega_\alpha   -\omega_j ) ( \omega_\beta   -\omega_j -\omega_k ) } 
- \frac{\lambda _{j\beta k}\lambda^* _{\beta \alpha k}\: a_k^\dagger\: a_k\: \sigma_{\beta \beta }}
{( \omega_\alpha   -\omega_j ) ( \omega_\beta   -\omega_j +\omega_k ) } 
\nonumber \\ 
\lo- \frac{\lambda _{\beta \alpha }\lambda^* _{j\beta }\: b\: b^\dagger\: \sigma_{\beta \beta }}
 {( \omega_\alpha   -\omega_j ) ( \omega_\beta   -\omega_j -\bar{\omega} ) } 
- \frac{\lambda^* _{\beta \alpha }\lambda^* _{j\beta }\: b^\dagger\: b^\dagger\: \sigma_{\beta \beta }}
{( \omega_\alpha   -\omega_j -2  \bar{\omega} ) ( \omega_\beta   -\omega_j 
-\bar{\omega} ) } 
\nonumber \\ 
\lo- \frac{\lambda^* _{\beta \alpha }\lambda^* _{j\beta k}\: b^\dagger\: a_k^\dagger\:
  \sigma_{\beta \beta }}{( \omega_\beta   -\omega_j -\omega_k ) ( \omega_\alpha   -\omega_j
   -\omega_k -\bar{\omega} ) } 
- \frac{\lambda^* _{\beta \alpha k}\lambda^* _{j\beta }\: a_k^\dagger\: 
b^\dagger\: \sigma_{\beta \beta }}{( \omega_\beta   -\omega_j -\bar{\omega} )
 ( \omega_\alpha   -\omega_j -\omega_k -\bar{\omega} ) } \nonumber \\ 
\lo- \frac{\lambda _{j\beta k}\lambda^* _{\beta \alpha }\: b^\dagger\: a_k\: \sigma_{\beta \beta }}
 {( \omega_\beta   -\omega_j +\omega_k ) 
 ( \omega_\alpha   -\omega_j +\omega_k -\bar{\omega} ) } 
- \frac{\lambda _{\beta \alpha k}\lambda^* _{j\beta }\: a_k\: b^\dagger\: \sigma_{\beta \beta }}
{( \omega_\beta   -\omega_j -\bar{\omega} ) 
( \omega_\alpha   -\omega_j +\omega_k -\bar{\omega} ) } 
\nonumber \\ 
\lo- \frac{\lambda _{j\beta }\lambda^* _{\beta \alpha }\: b^\dagger\: b\: \sigma_{\beta \beta }}
 {( \omega_\alpha   -\omega_j ) ( \omega_\beta   -\omega_j +\bar{\omega} ) } 
- \frac{\lambda _{\beta \alpha }\lambda^* _{j\beta k}\: b\: a_k^\dagger\: \sigma_{\beta \beta }}
{( \omega_\beta   -\omega_j -\omega_k )
 ( \omega_\alpha   -\omega_j -\omega_k +\bar{\omega} ) } 
\nonumber \\ 
\lo- \frac{\lambda _{j\beta }\lambda^* _{\beta \alpha k}\: a_k^\dagger\: b\: \sigma_{\beta \beta }}
 {( \omega_\beta   -\omega_j +\bar{\omega} )
  ( \omega_\alpha   -\omega_j -\omega_k +\bar{\omega} ) } 
- \frac{\lambda _{\beta \alpha }\lambda _{j\beta k}\: b\: a_k\: \sigma_{\beta \beta }}
{( \omega_\beta   -\omega_j +\omega_k )
 ( \omega_\alpha   -\omega_j +\omega_k +\bar{\omega} ) } 
\nonumber \\ 
\lo- \frac{\lambda _{\beta \alpha k}\lambda _{j\beta }\: a_k\: b\: \sigma_{\beta \beta }}
 {( \omega_\beta   -\omega_j +\bar{\omega} ) 
 ( \omega_\alpha   -\omega_j +\omega_k +\bar{\omega} ) } 
- \frac{\lambda _{\beta \alpha }\lambda _{j\beta }\: b\: b\: \sigma_{\beta \beta }}
{( \omega_\beta   -\omega_j +\bar{\omega} )
 ( \omega_\alpha   -\omega_j +2  \bar{\omega} ) } \: , 
\end{eqnarray}
\begin{eqnarray}
\fl A_{1j}^{12}(n) = 
 \frac{\lambda^* _{jnk}\lambda^* _{n2}\: b^\dagger\: a_k^\dagger\: \sigma_{12}}
 {( \omega_1  -\omega_j ) ( \omega_1  -\omega_n +\omega_k ) } 
+ \frac{\lambda _{jnk}\lambda^* _{n2}\: b^\dagger\: a_k\: \sigma_{12}}
{( \omega_1  -\omega_n +\omega_k ) 
( \omega_1  -\omega_j +2 \omega_k ) } 
\nonumber \\ 
\lo+ \frac{\lambda^* _{jn}\lambda^* _{n2}\: b^\dagger\: 
 b^\dagger\: \sigma_{12}}
 {( \omega_1  -\omega_n +\omega_k ) 
 ( \omega_1  -\omega_j +\omega_k -\bar{\omega} ) } 
+ \frac{\lambda^* _{jn}\lambda^* _{n2k}\: a_k^\dagger\: 
b^\dagger\: \sigma_{12}}{( \omega_1  -\omega_j )
 ( \omega_1  -\omega_n +\bar{\omega} ) } 
\nonumber \\ 
\lo+ \frac{\lambda _{jn}\lambda^* _{n2}\: b^\dagger\: b\: \sigma_{12}}
 {( \omega_1  -\omega_n +\omega_k ) 
 ( \omega_1  -\omega_j +\omega_k +\bar{\omega} ) } 
+ \frac{\lambda _{jnk}\lambda^* _{n2k}\: a_k^\dagger\: a_k\: \sigma_{12}}
{( \omega_1  -\omega_n +\bar{\omega} ) 
( \omega_1  -\omega_j +\omega_k +\bar{\omega} ) } 
\nonumber \\ 
\lo+ \frac{\lambda _{n2k}\lambda^* _{jn}\: a_k\: b^\dagger\: \sigma_{12}}
 {( \omega_1  -\omega_j +2 \omega_k ) 
 ( \omega_1  -\omega_n +2 \omega_k +\bar{\omega} ) } 
+ \frac{\lambda _{n2k}\lambda^* _{jnk}\: a_k\: a_k^\dagger\: \sigma_{12}}
{( \omega_1  -\omega_j +\omega_k +\bar{\omega} ) 
( \omega_1  -\omega_n +2 \omega_k +\bar{\omega} ) } 
\nonumber \\ 
\lo+ \frac{\lambda _{jn}\lambda^* _{n2k}\: a_k^\dagger\: b\: \sigma_{12}}
 {( \omega_1  -\omega_n +\bar{\omega} )
  ( \omega_1  -\omega_j +2 \bar{\omega} ) } 
+ \frac{\lambda _{n2}\lambda^* _{jn}\: b\: b^\dagger\: \sigma_{12}}
{( \omega_1  -\omega_j +\omega_k +\bar{\omega} ) 
( \omega_1  -\omega_n +\omega_k +2 \bar{\omega} ) } 
\nonumber \\ 
\lo+ \frac{\lambda _{n2}\lambda^* _{jnk}\: b\: a_k^\dagger\: \sigma_{12}}
 {( \omega_1  -\omega_j +2 \bar{\omega} )
  ( \omega_1  -\omega_n +\omega_k +2 \bar{\omega} ) } 
+ \frac{\lambda _{jn}\lambda _{n2k}\: a_k\: b\: \sigma_{12}}
{( \omega_1  -\omega_n +2 \omega_k +\bar{\omega} ) 
( \omega_1  -\omega_j +2 \omega_k +2 \bar{\omega} ) } 
\nonumber \\ 
\lo+ \frac{\lambda _{jnk}\lambda _{n2}\: b\: a_k\: \sigma_{12}}
 {( \omega_1  -\omega_n +\omega_k +2 \bar{\omega} )
  ( \omega_1  -\omega_j +2 \omega_k +2 \bar{\omega} ) } 
\nonumber \\ 
\lo+ \frac{\lambda _{jn}\lambda _{n2}\: b\: b\: \sigma_{12}}
{( \omega_1  -\omega_n +\omega_k +2 \bar{\omega} ) 
( \omega_1  -\omega_j +\omega_k +3 \bar{\omega} ) } \: , 
\end{eqnarray}
\begin{eqnarray}
\fl A_{1j}^{21} =
- \frac{\lambda _{21k}\lambda _{j1}\: a_k\: b\: \sigma_{21}}
{( \omega_1  -\omega_j ) ( \omega_2  -\omega_j -\omega_k ) }
- \frac{\lambda _{j1}\lambda^* _{21k}\: a_k^\dagger\: b\: \sigma_{21}}
{( \omega_1  -\omega_j -2 \omega_k ) ( \omega_2  -\omega_j -\omega_k ) }
\nonumber \\
\lo- \frac{\lambda^* _{21}\lambda^* _{j1}\: b^\dagger\: b^\dagger\: \sigma_{21}}
 {( \omega_1  -\omega_j -\omega_k -3 \bar{\omega} )
  ( \omega_2  -\omega_j -\omega_k -2 \bar{\omega} ) }
- \frac{\lambda _{21k}\lambda^* _{j1}\: a_k\: b^\dagger\: \sigma_{21}}
{( \omega_1  -\omega_j -2 \bar{\omega} ) 
( \omega_2  -\omega_j -\omega_k -2 \bar{\omega} ) } 
\nonumber \\
\lo- \frac{\lambda^* _{21k}\lambda^* _{j1}\: a_k^\dagger\: b^\dagger\: \sigma_{21}}
 {( \omega_1  -\omega_j -2 \omega_k -2 \bar{\omega} ) 
 ( \omega_2  -\omega_j -\omega_k -2 \bar{\omega} ) } 
- \frac{\lambda _{21}\lambda _{j1k}\: b\: a_k\: \sigma_{21}}
{( \omega_1  -\omega_j ) ( \omega_2  -\omega_j -\bar{\omega} ) } 
\nonumber \\
\lo- \frac{\lambda _{j1k}\lambda^* _{21}\: b^\dagger\: a_k\: \sigma_{21}}
 {( \omega_1  -\omega_j -2 \bar{\omega} ) ( \omega_2  -\omega_j -\bar{\omega} ) }
- \frac{\lambda _{21}\lambda^* _{j1k}\: b\: a_k^\dagger\: \sigma_{21}}
{( \omega_1  -\omega_j -2 \omega_k ) ( \omega_2  -\omega_j -2 \omega_k -\bar{\omega} ) } 
\nonumber \\
\lo- \frac{\lambda^* _{21}\lambda^* _{j1k}\: b^\dagger\: a_k^\dagger\: \sigma_{21}}
 {( \omega_1  -\omega_j -2 \omega_k -2 \bar{\omega} ) 
 ( \omega_2  -\omega_j -2 \omega_k -\bar{\omega} ) }
- \frac{\lambda _{j1}\lambda^* _{21}\: b^\dagger\: b\: \sigma_{21}}
{( \omega_2  -\omega_j -\omega_k ) ( \omega_1  -\omega_j -\omega_k -\bar{\omega} ) } 
\nonumber \\
\lo- \frac{\lambda _{21}\lambda^* _{j1}\: b\: b^\dagger\: \sigma_{21}}
 {( \omega_2  -\omega_j -\omega_k -2 \bar{\omega} )
  ( \omega_1  -\omega_j -\omega_k -\bar{\omega} ) }
- \frac{\lambda _{j1k}\lambda^* _{21k}\: a_k^\dagger\: a_k\: \sigma_{21}}
{( \omega_2  -\omega_j -\bar{\omega} ) ( \omega_1  -\omega_j -\omega_k -\bar{\omega} ) } 
\nonumber \\
\lo- \frac{\lambda _{21k}\lambda^* _{j1k}\: a_k\: a_k^\dagger\: \sigma_{21}}
 {( \omega_2  -\omega_j -2 \omega_k -\bar{\omega} ) 
 ( \omega_1  -\omega_j -\omega_k -\bar{\omega} ) }
\nonumber \\
\lo- \frac{\lambda _{21}\lambda _{j1}\: b\: b\: \sigma_{21}}
{( \omega_2  -\omega_j -\omega_k ) ( \omega_1  -\omega_j -\omega_k +\bar{\omega} ) }\, ,
\end{eqnarray}
\begin{eqnarray}
\fl A_{2j}^{11}= {\mathcal B}(2,1)\, ,
\end{eqnarray}
\begin{eqnarray}
\fl A_{2j}^{22}(n)= \mathcal{A}(2,n)\, ,
\end{eqnarray}
\begin{eqnarray}
\fl A_{2j}^{12}= 
- \frac{\lambda^* _{12k}\lambda^* _{j2}\: a_k^\dagger\: 
b^\dagger\: \sigma_{12}}{( \omega_2  -\omega_j ) 
( \omega_1  -\omega_j +\omega_k ) } 
- \frac{\lambda _{12k}\lambda^* _{j2}\: a_k\: b^\dagger\: \sigma_{12}}
{( \omega_1  -\omega_j +\omega_k ) 
( \omega_2  -\omega_j +2 \omega_k ) } 
\nonumber \\ 
\lo- \frac{\lambda^* _{12}\lambda^* _{j2}\: b^\dagger\: 
 b^\dagger\: \sigma_{12}}{( \omega_1  -\omega_j +\omega_k )
  ( \omega_2  -\omega_j +\omega_k -\bar{\omega} ) } 
- \frac{\lambda^* _{12}\lambda^* _{j2k}\: b^\dagger\: a_k^\dagger\:
 \sigma_{12}}{( \omega_2  -\omega_j ) 
 ( \omega_1  -\omega_j +\bar{\omega} ) } 
\nonumber \\ 
\lo- \frac{\lambda _{12}\lambda^* _{j2}\: b\: b^\dagger\: \sigma_{12}}
 {( \omega_1  -\omega_j +\omega_k )
  ( \omega_2  -\omega_j +\omega_k +\bar{\omega} ) } 
- \frac{\lambda _{12k}\lambda^* _{j2k}\: a_k\: a_k^\dagger\: \sigma_{12}}
{( \omega_1  -\omega_j +\bar{\omega} )
 ( \omega_2  -\omega_j +\omega_k +\bar{\omega} ) } 
\nonumber \\ 
\lo- \frac{\lambda _{j2k}\lambda^* _{12}\: b^\dagger\: a_k\: \sigma_{12}}
 {( \omega_2  -\omega_j +2 \omega_k ) 
 ( \omega_1  -\omega_j +2 \omega_k +\bar{\omega} ) } 
- \frac{\lambda _{j2k}\lambda^* _{12k}\: a_k^\dagger\: a_k\: \sigma_{12}}
{( \omega_2  -\omega_j +\omega_k +\bar{\omega} ) 
( \omega_1  -\omega_j +2 \omega_k +\bar{\omega} ) } 
\nonumber \\ 
\lo- \frac{\lambda _{12}\lambda^* _{j2k}\: b\: a_k^\dagger\: \sigma_{12}}
 {( \omega_1  -\omega_j +\bar{\omega} ) 
 ( \omega_2  -\omega_j +2 \bar{\omega} ) } 
- \frac{\lambda _{j2}\lambda^* _{12}\: b^\dagger\: b\: \sigma_{12}}
{( \omega_2  -\omega_j +\omega_k +\bar{\omega} )
 ( \omega_1  -\omega_j +\omega_k +2 \bar{\omega} ) } 
\nonumber \\ 
\lo- \frac{\lambda _{j2}\lambda^* _{12k}\: a_k^\dagger\: b\: \sigma_{12}}
 {( \omega_2  -\omega_j +2 \bar{\omega} ) 
 ( \omega_1  -\omega_j +\omega_k +2 \bar{\omega} ) } 
- \frac{\lambda _{12}\lambda _{j2k}\: b\: a_k\: \sigma_{12}}
{( \omega_1  -\omega_j +2 \omega_k +\bar{\omega} ) 
( \omega_2  -\omega_j +2 \omega_k +2 \bar{\omega} ) } 
\nonumber \\ 
\lo- \frac{\lambda _{12k}\lambda _{j2}\: a_k\: b\: \sigma_{12}}
 {( \omega_1  -\omega_j +\omega_k +2 \bar{\omega} ) 
 ( \omega_2  -\omega_j +2 \omega_k +2 \bar{\omega} ) } 
\nonumber \\ 
\lo- \frac{\lambda _{12}\lambda _{j2}\: b\: b\: \sigma_{12}}
{( \omega_1  -\omega_j +\omega_k +2 \bar{\omega} ) 
( \omega_2  -\omega_j +\omega_k +3 \bar{\omega} ) } \: ,
\end{eqnarray}
\begin{eqnarray}
\fl A_{2j}^{21}(n)= 
 \frac{\lambda _{jnk}\lambda _{n1}\: b\: a_k\: \sigma_{21}}
 {( \omega_2  -\omega_j ) ( \omega_2  -\omega_n -\omega_k ) } 
+ \frac{\lambda _{n1}\lambda^* _{jnk}\: b\: a_k^\dagger\: \sigma_{21}}
{( \omega_2  -\omega_j -2 \omega_k ) 
( \omega_2  -\omega_n -\omega_k ) } 
\nonumber \\ 
\lo+ \frac{\lambda^* _{jn}\lambda^* _{n1}\: b^\dagger\: b^\dagger\:
  \sigma_{21}}{( \omega_2  -\omega_j -\omega_k -3 \bar{\omega} )
   ( \omega_2  -\omega_n -\omega_k -2 \bar{\omega} ) } 
+ \frac{\lambda _{jnk}\lambda^* _{n1}\: b^\dagger\: a_k\: \sigma_{21}}
{( \omega_2  -\omega_j -2 \bar{\omega} ) 
( \omega_2  -\omega_n -\omega_k -2 \bar{\omega} ) } 
\nonumber \\ 
\lo+ \frac{\lambda^* _{jnk}\lambda^* _{n1}\: b^\dagger\: 
 a_k^\dagger\: \sigma_{21}}
 {( \omega_2  -\omega_j -2 \omega_k -2 \bar{\omega} )
  ( \omega_2  -\omega_n -\omega_k -2 \bar{\omega} ) } 
+ \frac{\lambda _{jn}\lambda _{n1k}\: a_k\: b\: \sigma_{21}}
{( \omega_2  -\omega_j ) ( \omega_2  -\omega_n -\bar{\omega} ) } 
\nonumber \\ 
\lo+ \frac{\lambda _{n1k}\lambda^* _{jn}\: a_k\: b^\dagger\: \sigma_{21}}
 {( \omega_2  -\omega_j -2 \bar{\omega} ) 
 ( \omega_2  -\omega_n -\bar{\omega} ) } 
+ \frac{\lambda _{jn}\lambda^* _{n1k}\: a_k^\dagger\: b\: \sigma_{21}}
{( \omega_2  -\omega_j -2 \omega_k )
 ( \omega_2  -\omega_n -2 \omega_k -\bar{\omega} ) } 
\nonumber \\ 
\lo+ \frac{\lambda^* _{jn}\lambda^* _{n1k}\: a_k^\dagger\: 
 b^\dagger\: \sigma_{21}}{( \omega_2  -\omega_j -2 \omega_k -2 \bar{\omega} ) 
 ( \omega_2  -\omega_n -2 \omega_k -\bar{\omega} ) } 
+ \frac{\lambda _{n1}\lambda^* _{jn}\: b\: b^\dagger\: \sigma_{21}}
{( \omega_2  -\omega_n -\omega_k ) 
( \omega_2  -\omega_j -\omega_k -\bar{\omega} ) } 
\nonumber \\ 
\lo+ \frac{\lambda _{jn}\lambda^* _{n1}\: b^\dagger\: b\: \sigma_{21}}
 {( \omega_2  -\omega_n -\omega_k -2 \bar{\omega} ) 
 ( \omega_2  -\omega_j -\omega_k -\bar{\omega} ) } 
+ \frac{\lambda _{n1k}\lambda^* _{jnk}\: a_k\: a_k^\dagger\: \sigma_{21}}
{( \omega_2  -\omega_n -\bar{\omega} ) 
( \omega_2  -\omega_j -\omega_k -\bar{\omega} ) } 
\nonumber \\ 
\lo+ \frac{\lambda _{jnk}\lambda^* _{n1k}\: a_k^\dagger\: a_k\: \sigma_{21}}
 {( \omega_2  -\omega_n -2 \omega_k -\bar{\omega} )
  ( \omega_2  -\omega_j -\omega_k -\bar{\omega} ) } 
\nonumber \\ 
\lo+ \frac{\lambda _{jn}\lambda _{n1}\: b\: b\: \sigma_{21}}
{( \omega_2  -\omega_n -\omega_k ) 
( \omega_2  -\omega_j -\omega_k +\bar{\omega} ) } 
\: .
\end{eqnarray}
%

\end{document}